\documentclass[12pt]{article}

\usepackage[top=80pt,bottom=85pt,left=58pt,right=58pt]{geometry}
\usepackage{amssymb,amsmath,xypic}
\usepackage{slashed}
\usepackage{graphicx,color,xcolor,subfigure}
\usepackage{float}
\usepackage{tensor}
\usepackage{sectsty}
\usepackage{cite}
\usepackage{bbm}
\usepackage[debug,pageanchor=false]{hyperref}
\definecolor{link}{rgb}{.8,.15,.1}
\hypersetup{colorlinks=true,linkcolor=link,citecolor=link,urlcolor=link,linktocpage}

\usepackage{tikz}
\makeatletter
\@addtoreset{equation}{section}
\makeatother

\renewcommand{\theequation}{\thesection.\arabic{equation}}

\newcommand{\mrm}[1]{\mbox{$\mathrm{#1}$}}
\newcommand{\beq}{\begin{equation}}
\newcommand{\eeq}{\end{equation}}
\newcommand{\bea}{\begin{eqnarray}}
\newcommand{\eea}{\end{eqnarray}}
\newcommand{\nn}{\nonumber}

\newcommand{\eq}{\begin{equation}}
\newcommand{\feq}{\end{equation}}
\newcommand{\eqn}{\begin{eqnarray}}
\newcommand{\feqn}{\end{eqnarray}}

\begin{document}

\begin{titlepage}

\begin{center}

\vskip .5in 
\noindent

{\Large \bf{On $\text{AdS}_2\times \text{S}^7$, its $\mathbb{Z}_k$ orbifold and their dual quantum mechanics}}

\bigskip\medskip

Yolanda Lozano$^{a,b}$\footnote{ylozano@uniovi.es}, Niall T. Macpherson$^{a,b}$\footnote{macphersonniall@uniovi.es}, Achilleas Passias$^{c}$\footnote{achilleas.passias@uoi.gr},  \\

\bigskip\bigskip\bigskip
{\small 

a: Department of Physics, University of Oviedo,
Avda. Federico Garcia Lorca s/n, 33007 Oviedo}

\medskip
{\small and}

\medskip
{\small 

b: Instituto Universitario de Ciencias y Tecnolog\'ias Espaciales de Asturias (ICTEA),\\
Calle de la Independencia 13, 33004 Oviedo, Spain}

\bigskip\medskip
{\small 

c: Department of Physics, University of Ioannina, GR45110, Ioannina, Greece }

\bigskip\medskip

{\bf Abstract }\\[2mm]
\end{center}   	
We consider a previously constructed class of massive Type IIA AdS$_2\times$S$^7\times I$ solutions with OSp$(8|2)$ symmetry, as well as OSp$(6|2)$-symmetric ones, by replacing the S$^7$ with the orbifold S$^7/\mathbb{Z}_k$. In both cases we construct global solutions for which the interval $I$ is bounded between physical singularities, by allowing D8-branes transverse to $I$. We also generate a new class of Type IIB AdS$_2\times \mathbb{CP}^3\times\text{S}^1\times I$ solutions by T-duality and establish a chain of dualities that maps the massless limit of these classes to AdS$_4/\mathbb{Z}_{k'}\times\text{S}^7/\mathbb{Z}_k$, thus identifying the brane configurations yielding these solutions. We propose that the ${\cal N}=8$ solutions are dual to a theory living on a  D0-F1-D8 brane intersection which has a description in terms of disconnected quivers and similarly for the ${\cal N}=6$ solutions.

	


\end{titlepage}

\setcounter{footnote}{0}

\tableofcontents

\setcounter{footnote}{0}
\renewcommand{\theequation}{{\rm\thesection.\arabic{equation}}}

\section{Introduction}

Owning primarily to the AdS/CFT correspondence, there is by now a large amount of work on anti-de Sitter solutions of string/M-theory in diverse dimensions. The case of AdS$_2$ stands out, as the near-horizon of extremal black holes and due to the non-connectedness of its boundary, among other conceptual challenges that the holographic duality between AdS$_2$ and its dual superconformal quantum mechanics (SCQM) possesses \cite{Strominger:1998yg, Maldacena:1998uz}. The high dimensionality of the internal space of AdS$_2$ solutions of string/M-theory allows for a large landscape of solutions; a non-exhaustive list of recent explorations is \cite{Hong:2019wyi, Dibitetto:2019nyz, Corbino:2020lzq, Lozano:2020txg, Lozano:2020sae, Lozano:2021rmk, Ramirez:2021tkd, Lozano:2021fkk, Lozano:2022vsv, Lozano:2022swp, Conti:2023naw, Conti:2023rul, Legramandi:2023fjr}.

The AdS$_2$ spacetime emerges in the near-horizon limit of extremal black holes and several physical properties of a black hole depend only on its near-horizon geometry, like for example the Bekenstein--Hawking entropy. Seminally, the embedding of the five-dimensional, asymptotically flat, extremal Reissner--Nordstrom black hole in type IIB string theory, enabled Strominger and Vafa to count the microstates accounting for its entropy in \cite{Strominger:1996sh}. The AdS/CFT correspondence provides a window into the microstates for asymptotically AdS$_4$ black holes preserving ${\cal N}=2$ supersymmetry. Here one computes a topologically twisted index in the dual CFT$_3$ \cite{Benini:2015eyy,Benini:2016rke} that can be compared against the black hole entropy. Interestingly, such CFT methods suggest there are many more such black holes than currently constructed.

AdS$_2$ solutions also emerge in several higher-dimensional AdS/CFT contexts. Examples include: scenarios with branes wrapped on Riemann or higher-dimensional surfaces described by compactified CFTs, such as \cite{Faedo:2021nub,Couzens:2022lvg,Faedo:2024upq}. Janus and Hades solutions with global AdS$_2$ factors which merge with some portion of their internal space to form higher-dimensional AdS factors at certain loci. Such solutions are dual to interfaces between higher-dimensional CFTs; see for instance \cite{Bak:2007jm,Chen:2021mtn,Gutperle:2024yiz}. AdS$_2$ solutions also  appear in the holographic  description of defects in higher-dimensional CFTs as in \cite{DHoker:2007mci,Chen:2019qib,Arav:2024wyg}. In contrast, the present work addresses AdS$_2$/SCQM correspondence without a higher-dimensional origin.

A class of $\text{AdS}_2\times \text{S}^7\times I$ solutions in massive Type IIA supergravity, preserving $\mathcal{N}=8$ supersymmetry, was constructed in \cite{Dibitetto:2018gbk}. They arise as the near-horizon limit of a $\frac{1}{4}$-BPS D0-F1-D8 brane configuration found in \cite{Imamura:2001cr}. In the case of zero Romans mass, the corresponding solution was shown later to arise from the $\text{AdS}_4\times \text{S}^7$ M-theory solution, through reduction along the Hopf fibre of the $\text{AdS}_3$ contained in $\text{AdS}_4$ \cite{Dibitetto:2019nyz}. 

In this paper we revisit the solutions of \cite{Dibitetto:2018gbk} and extend them in multiple ways. 
We mod the S$^7$ by $\mathbb{Z}_k$, thus breaking the supersymmetry to $\mathcal{N}=6$. While in \cite{Dibitetto:2018gbk} the $\text{AdS}_2\times \text{S}^7\times I$ solutions where non-compact with a semi-infinite interval $I$, in the present work we construct compact solutions by including D8-brane
sources along the interior of the interval. This allows one to glue finite portions of the semi-infinite solutions together  and thereby arrive at a  compact internal space. At the end-points of the now finite interval $I$, it is possible to have one of three distinct behaviours: a $\mathbb{R}^8/\mathbb{Z}_k$ orbifold singularity, an OF1-plane (S-dual to an O1-plane) or a D8/O8 singularity. It is actually possible to place an arbitrary number of D8-branes in the interior of $I$, so one can construct infinitely many solutions in this fashion.

\begin{figure}[http!]
	\begin{center}
		\scalebox{1.0}[1.0]{ \xymatrix@C-0pc { 
				\text{  } & *+[F-,]{\textrm{AdS}_4/\mathbb{Z}_{k'}\times \text{S}^7/\mathbb{Z}_k} \ar[dl]_{\text{S}^1/\mathbb{Z}_{k'}} \ar[dr]^{\text{S}^1/\mathbb{Z}_k} & \text{  }    \\
		*+[F-,]{\textrm{AdS}_2\times \text{S}^7/\mathbb{Z}_k \times I}\ar@{<->}[d]^{\mathrm{T}_{\text{S}^1/\mathbb{Z}_k}} &  \text{  }  & 
		*+[F-,]{ \textrm{AdS}_4/\mathbb{Z}_{k'} \times \mathbb{CP}^3}\ar@{<->}[d]^{\mathrm{T}_{\text{S}^1/\mathbb{Z}_{k'}}}\\
		*+[F-,]{\textrm{AdS}_2\times \mathbb{CP}^3  \times  \text{S}^1/\mathbb{Z}_k\times I} &  \ar[l] \text{S} \ar[r] & 
		*+[F-,]{\textrm{AdS}_2\times \mathbb{CP}^3\times \text{S}^1/\mathbb{Z}_{k'}\times I}
		}}
	\end{center}
	\caption{Web of dualities that relate the new AdS$_2$ solution constructed in section \ref{sec:N=6} (on the left-hand side) with the solution constructed in  \cite{Conti:2023rul} corresponding to zero Romans mass (on the right-hand side).}\label{fig:dualities} 
\end{figure}
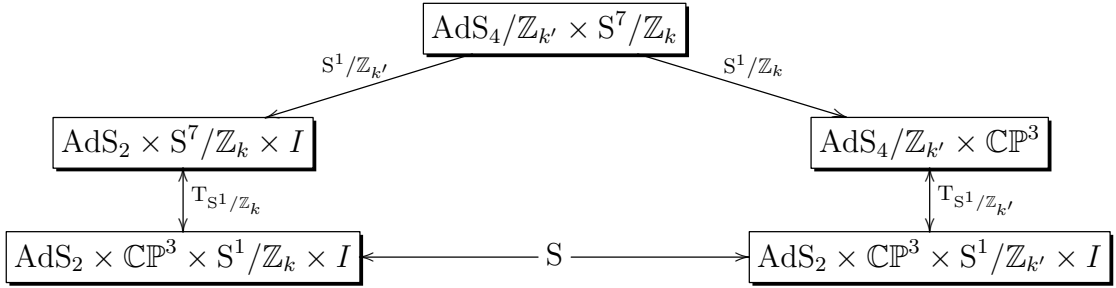

We further uncover a web of AdS$_2$ solutions by applying T- and S-dualities. We perform two T-duality transformations that map the KK-monopoles which feature in the Type IIA geometry onto NS5-branes and lead to a new class of $\text{AdS}_2$ solutions in Type IIB. These solutions are warped products of $\text{AdS}_2\times \mathbb{CP}^3\times \text{S}^1$ and an interval, preserving $\mathcal{N}=6$ supersymmetry. By taking the limit of vanishing axion field, we can perform an S-duality transformation, and thereby, establish a relation with the class of $\text{AdS}_2\times \mathbb{CP}^3\times \text{S}^1\times I$ solutions constructed recently in \cite{Conti:2023rul}\footnote{These were constructed by T-dualising the $\text{AdS}_3\times \mathbb{CP}^3\times I$ solutions of massive Type IIA supergravity found in \cite{Macpherson:2023cbl}.}. In this manner, we are able to relate both types of solutions to the $\text{AdS}_4/\mathbb{Z}_{k'}\times \text{S}^7/\mathbb{Z}_k$ one in eleven dimensions and identify the brane intersection that underlies them. The web of AdS$_2$ solutions is summarised in Figure \ref{fig:dualities}.
Having identified the underlying brane set-up we extend it to the massive case and proceed with the construction of the quantum mechanics that emerges through the quantisation of the open strings. 

Our analysis of the brane intersection underlying the ${\cal N}=8$ AdS$_2$ solutions leads us to propose that their dual SCQM is realised at the IR fixed point of the theory living on a D0-F1-D8 system. This has a description in terms of disconnected quivers involving ${\cal N} = 8$ vector, ${\cal N} = 8$
hyper and  ${\cal N} = 2$ Fermi multiplets. It is also possible to interpret the ${\cal N}=8$ AdS$_2$ solutions as realising backreacted baryon vertices for the field theories living on D8-branes. The ${\cal N}=6$ AdS$_2$ solutions are more complicated, but by making a connection to ABJM we are still able to make progress. When the Romans mass is zero these solutions can be realised by adding momentum to M2-branes backreacted on $\mathbb{C}^4/\mathbb{Z}_k$. This suggests that the dual SCQM for generic Romans mass should be the IR fixed point of the ${\cal N}=3$ quantum mechanics living on the intersection of the Type IIA reduction of this theory, namely, the ABJM Type IIA brane set-up with extra waves, and D8-branes. We make a proposal for describing this in terms of more general disconnected quivers that reproduce the ${\cal N}=8$ analogue in the appropriate limit. Again, there is an interpretation in terms of baryon vertices. 

The paper is organised as follows. In section \ref{sec:N=6IIA} we present the $\text{AdS}_2\times \text{S}^7\times I$ solutions of Type IIA and extend their global structure. In section \ref{sec:N=6} we construct a new  $\text{AdS}_2\times \mathbb{CP}^3\times \text{S}^1\times I$ class in type IIB that preserves ${\cal N}=6$ supersymmetry via dualities and establish the connection to $\text{AdS}_4/\mathbb{Z}_{k'}\times \text{S}^7/\mathbb{Z}_k$. Sections \ref{sec:N=8FT} and \ref{sec:N=6FT} contain the field theory interpretation for $\mathcal{N}=8$ and $\mathcal{N}=6$ supersymmetric theories respectively.

\section{$\text{AdS}_2\times\text{S}^7/\mathbb{Z}_k \times I$ with $\mathcal{N}=6$ in massive Type IIA}
\label{sec:N=6IIA}

A class of AdS$_2\times\text{S}^7 \times I$ solutions to massive Type IIA with OSp$(8|2)$ superconformal symmetry was constructed in \cite{Dibitetto:2018gbk}. These solutions preserve $\mathcal{N}=8$ supersymmetry locally, with the SO(8) R-symmetry realised geometrically on the $\text{S}^7$. By expressing the 7-sphere as a U(1) fibration over $\mathbb{CP}^3$ and then taking a $\mathbb{Z}_k$ orbifold of the fibre direction one preserves ${\cal N}=6$ supersymmetry globally. The resulting solution has the following  form
\begin{eqnarray}\label{solution}
&&\frac{ds^2}{L^2}=\frac{h^{3/2}\sqrt{h''}}{8(2hh''-h'^2)}ds^2(\text{AdS}_2)+\frac18 \sqrt{\frac{h''}{h}}dr^2+\sqrt{\frac{h}{h''}}ds^2(\text{S}^7/\mathbb{Z}_k)\label{metric}\\
&&e^{\Phi}=\frac{c_0L^3}{\sqrt{2hh''-h'^2}}\Bigl(\frac{h}{h''}\Bigr)^{3/4}, \qquad B_2=-\frac{L^2}{8\sqrt{2}} \Bigl(\frac{hh'}{2hh''-h'^2}+r-l\Bigr)\text{vol}(\text{AdS}_2)\label{NSNS1}\\
&&F_0=\frac{4}{c_0L^4}h'''\label{Bianchi}\\
&&\hat{F}_2=\frac{1}{2\sqrt{2}c_0L^2}\Bigl((r-l)h'''-h''\Bigr)\text{vol}(\text{AdS}_2)\label{RR1}
\end{eqnarray}
where $r$ parametrises the interval, $h(r)$ satisfies \eqref{Bianchi},
with $F_0$ the Romans mass, and $\hat{F}_2$ is the Page flux $F_2-F_0 B_2$.  As noted in \cite{Dibitetto:2018gbk} the AdS$_2\times\text{S}^7 \times I$ solutions can be obtained from the $\text{AdS}_7\times \text{S}^2\times I$ solutions of massive Type IIA supergravity constructed in \cite{Apruzzi:2013yva} through a double analytical continuation. When the S$^7$ is orbifolded by $\mathbb{Z}
_k$ the solutions are obtained from $\text{AdS}_7/\mathbb{Z}_k \times \text{S}^2\times I$. As for the solutions in  \cite{Apruzzi:2013yva} the $r$-direction must be divided in intervals to enforce that $B_2$ lies in the fundamental region,
\begin{equation}\label{fundamental}
\frac{1}{4\pi^2}\bigg|\int_{\text{AdS}_2}B_2\bigg|\in [0,1].
\end{equation}
We will choose the conventions 
\begin{equation}\label{conventions}
L^2=8\pi\sqrt{2},\qquad c_0=(16\pi)^{-1},\qquad \int_{\text{AdS}_2}\text{vol}(\text{AdS}_2)=\text{Vol}(\text{AdS}_2)=4\pi
\end{equation}
such that we can take $l$ to be an integer\footnote{This prescription is based on the analytic continuation relating AdS$_2$ to S$^2$.}. This is the gauge parameter of the large gauge transformation that must be performed in each $[l,l+1]$ interval to enforce the condition \eqref{fundamental}.
In this case this creates an F1-string stretched between $r=l$ and $r=l+1$, through an electric-magnetic analogue of the NS5-brane creation effect \cite{Hanany:1996ie}, that takes place for the solutions in  \cite{Apruzzi:2013yva}.

The most general solution to the equations of motion can then be constructed by glueing the local solutions defined in $r\in [l,l+1]$ with D8-branes, through an electric-magnetic analogue of the constructions in \cite{Cremonesi:2015bld} (see also \cite{Apruzzi:2017nck}). We choose for this purpose
\begin{equation}\label{hder}
h_l(r)=\frac{1}{6}\beta_l (r-l)^3+\frac12 \gamma_l (r-l)^2+\delta_l (r-l)+\mu_l \qquad \text{for} \qquad r\in [l,l+1],
\end{equation}
with $\beta_l$, $\gamma_l$, $\delta_l$ and $\mu_l$ constants that must be tuned between intervals such that metric and dilaton are continuous. $\beta_l$ and $\gamma_l$ can be related to the quantised charges of D0 and D8-branes through equations \eqref{RR1} and \eqref{Bianchi}. From the last equation we see that the number of D8-branes at each $[l,l+1]$ interval is given by
\begin{equation}\label{QD8}
Q_{{\rm D}8}^l=\beta_l,
\end{equation}
while from equation \eqref{RR1} we see that there are D0-branes at each $[l,l+1]$ interval carrying $\gamma_k$ units of electric charge, through the coupling\footnote{This gives actually $-\gamma_l$. We will absorb these signs through the paper by taking anti-branes when appropriate.}
\begin{equation}\label{QD0}
Q_{{\rm D}0}^l=\frac{1}{2\pi}\int_{\text{AdS}_2} \hat{F}_2=\gamma_l,
\end{equation}
where we have again regularised $\text{Vol}(\text{AdS}_2)=4\pi$. Note that imposing continuity of $h''$ across intervals one finds that the electric D0-brane charge in a given $[l,l+1]$ interval is determined from the D8-brane charges in the previous $[j,j+1]$ intervals, as
\begin{equation}\label{betagamma}
Q_{{\rm D}0}^l=\sum_{j=0}^{l-1}Q_{{\rm D}8}^j\, .
\end{equation}

Previous to the orbifolding, the resulting D0-D8-F1 brane system, depicted in Table \ref{F1-D0-D8}, preserves half-maximal supersymmetry in one dimension, that is, $\mathcal{N}=8$, with OSp$(8|2)$ superconformal symmetry and SO(8) R-symmetry, that is realised geometrically on the S$^7$ \cite{Dibitetto:2018gbk}.
\begin{table}[h]
\renewcommand{\arraystretch}{1}
\begin{center}
\scalebox{1}[1]{
\begin{tabular}{c| c cc  c c  c  c c c c}
 branes & $x^0$ & $x^1$ & $x^2$ & $x^3$ & $x^4$ & $x^5$ & $x^6$ & $x^7$ & $x^8$ & $x^9$ \\
\hline \hline
$\mrm{F}1$ & $\times$ & $\times$  & $-$ & $-$ & $-$ & $-$ & $-$ & $-$ & $-$ & $-$ \\
$\mrm{D}0$ & $\times$ & $-$  & $-$ & $-$ & $-$ & $-$ & $-$ & $-$ & $-$ & $-$\\
$\mrm{D}8$ & $\times$ & $-$ & $\times$ & $\times$ & $\times$ & $\times$ & $\times$ & $\times$ & $\times$ & $\times$ \\
\end{tabular}
}
\caption{Brane intersection associated to the $\text{AdS}_2\times \text{S}^7 \times I $ solutions in \cite{Dibitetto:2018gbk}.}
 \label{F1-D0-D8}
\end{center}
\end{table}
When the S$^7$ is modded out by $\mathbb{Z}_k$, in its parametrisation as a U(1) fibre over  $\mathbb{CP}^3$, 
\begin{equation}\label{tau-direction}
ds^2(\text{S}^7/\mathbb{Z}_k)=\Bigl(\frac{d\tau}{k}+{\cal A}\Bigr)^2+ds^2(\mathbb{CP}^3),
\end{equation}
with $d{\cal A}=2J$ and $J$ the K\"ahler form of the $\mathbb{CP}^3$, supersymmetry is reduced to $\mathcal{N}=6$, with OSp$(6|2)$ superconformal symmetry and SO(6) R-symmetry, that is realised geometrically on the $\mathbb{CP}^3$. This introduces KK-monopoles in the brane intersection, as we will further discuss in section \ref{sec:N=6}.

For AdS$_2$/SCQM purposes we are interested in the solutions within the class of this section with a bounded internal space.  That means that in addition to imposing  that the metric and dilaton are continuous across intervals, we must arrange for there to be a first and last interval where $r$ terminates in a physical way. To this end in the next section we study the possible physical boundary behaviours.

\subsection{Physical boundary behaviours}
In this section we review the possible boundary behaviours that \eqref{solution} can support. This mostly follows the results of \cite{Dibitetto:2018gbk}, though some claims are sharpened. We will make the assumption that  $F_0\neq 0$ throughout this section, the $F_0=0$ limit is a reduction of AdS$_4\times$S$^7$ which we consider independently in section \ref{massless-limit}.\\
~\\
We begin by observing that a metric with the correct signature in \eqref{solution}  requires that we impose
\begin{equation}
2hh''-h'^2 \geq 0. \label{eq:metricposs}
\end{equation}
On the other hand the special loci of the geometry where the warp factor vanishes or 
tends to infinity are defined by one of the 3 conditions 
\begin{itemize}
\item $h = 0$,
\item $h'' = 0$,
\item $2hh''-h'^2 = 0$. 
\end{itemize}
The bound \eqref{eq:metricposs} demands that if $h'$ tends to zero we must have that one of $h$
or $h''$ behaves likewise with the other remaining finite. This observation allows one to identify  the following special loci:
\begin{enumerate}
\item a double root of $h$,
\item a triple root of $h$,
\item a root of $h''$ and $h'$ (stationary point of inflection of $h$),
\item a root of $2hh''-h'^2$, while keeping $h''$ and $h'$ different from $0$,
\end{enumerate}
where we only need to consider a simple root of $2hh''-h'^2$ because the discriminant of $2hh''-h'^2$ is proportional to that of $h$, which implies that a multiple root of $2hh''-h'^2$ is also a multiple root of $h$, which we already considered. 

Let us now analyse the solution near the aforementioned loci:\\
~\\
Near a double root $r_0$ of $h$ one can show that the metric  tends to
\begin{equation}
\frac{ds^2}{L^2} \sim \frac{3}{2^3\sqrt{2}} \frac{h''(r_0)}{h'''} ds^2(\text{AdS}_2) + \frac{1}{\sqrt{2}}\left(d\varrho^2 + \varrho^2 ds^2(\text{S}^7/\mathbb{Z}_k)\right)
\end{equation}
where $\varrho^2 \equiv r-r_0$, and the dilaton stays finite. The internal space thus exhibits an $\mathbb{R}^8/\mathbb{Z}_k$ orbifold singularity at such a locus (unless $k=1$ in which case there is a regular zero).\\
~\\
Near a triple root $r_0$ of $h$ the metric becomes
\begin{equation}
\frac{ds^2}{L^2} \sim \frac{\sqrt{6}}{8} \left[ \frac{dr^2}{r-r_0} + (r-r_0) \left(\frac{1}{3} ds^2(\text{AdS}_2) + \frac{4}{3} ds^2(\text{S}^7/\mathbb{Z}_k) \right) \right],\label{eq:notphysical}
\end{equation}
which we do not recognise as physical.\\
~\\
Near a stationary point of inflection of $h$ the metric reads 
\begin{equation}
\frac{ds^2}{L^2} \sim (r - r_0)^{1/2} \frac{1}{8} \sqrt{\frac{h'''}{h(r_0)}} dr^2 + (r - r_0)^{-1/2} \sqrt{\frac{h(r_0)}{h'''}} \left( \frac{1}{16} ds^2(\text{AdS}_2) + ds^2(\text{S}^7/\mathbb{Z}_k) \right) .
\end{equation}
The dilaton on the other hand becomes
\beq
e^\Phi \sim  2^{-1/2} c_0 L^3 (h(r_0))^{1/4} (h''')^{-5/4} (r-r_0)^{-5/4}.
\eeq
We recognise the singular, yet physical, behaviour as associated to a O8/D8-brane system extended in AdS$_2\times$S$^7$.\\
~\\
The final case we must consider is a simple root of $2hh''-h'^2$, at such loci the metric at leading order becomes
\begin{equation}
\frac{ds^2}{L^2} \sim \frac{1}{16} \frac{\sqrt{h(r_0)h''(r_0)}}{h'''} (r-r_0)^{-1} ds^2(\text{AdS}_2) + \frac{1}{8}\sqrt{\frac{h''(r_0)}{h(r_0)}} dr^2 + \sqrt{\frac{h(r_0)}{h''(r_0)}} ds^2(\text{S}^7/\mathbb{Z}_k),
\end{equation}
while the  dilaton behaves as
\beq
e^\Phi \sim 2^{-1/2} c_0 L^3 (h(r_0))^{1/4} (h''(r_0))^{-3/4} (h''')^{-1/2} (r-r_0)^{-1/2}.
\eeq
This is the singularity associated to a localised OF1-plane extended in AdS$_2$ and backreacted on S$^7$ \footnote{This is a revised interpretation of the singularity compared to \cite{Dibitetto:2018gbk}.}. Such objects are to O1-planes as fundamental strings are to D1-branes, \textit{i.e} in Type IIB they are S-dual to each other. Like fundamental strings, OF1-planes can exist in both Type IIA and IIB.\\
~\\
In summary, we find that it is possible to bound the interval spanned by $r$ at some finite point $r=r_0$ in three distinct and physical ways: an $\mathbb{R}^8/\mathbb{Z}_k$ orbifold, a O8/D8 system or a localised OF1-plane. If one assumes these are placed at $r=0$, as one is free to use the translational invariance of the defining ODE to ensure, they  amount to tuning $h$ in the following ways
\begin{align}
\mathbb{R}^8/\mathbb{Z}_k&:~~~~h=(c_1+c_2 r) r^2,\nn\\[2mm]
\text{O8/D8}&:~~~~h=c_1+ c_2 r^3,\nn\\[2mm]
\text{OF1}&:~~~~h= \frac{c_1^2}{4 c_2}+c_1 r+c_2 r^2+c_3 r^3 ,\label{eq:boundarybehaviours}
\end{align}
for $c_i$ integration constants which must be tuned so as to satisfy \eqref{eq:metricposs}.

Having established that one can define local solutions for which the $r$ interval is bounded at one end, a reasonable question to ask is: what is the entire domain of $r$ for such a solution? An analysis of the roots of the quartic polynomial $2hh''-h'^2$ shows that it always has two real roots (simple or multiple), however \eqref{eq:metricposs} is not satisfied between these. Rather, since the coefficient of the $r^4$ term of $2hh''-h'^2$ is positive, we conclude that $2hh''-h'^2>0$ can only be ensured  in the range $r \in [r_0, \infty]$ or $r \in [- \infty,r_0]$ where $r_0$ is a root. The behaviour of the solution at infinity is
\begin{equation}
\label{asymptotic}
\frac{ds^2}{L^2} \sim \frac{\sqrt{6}}{8} \left[ \frac{dr^2}{r} + r \left(\frac{1}{3} ds^2(\text{AdS}_2) +  \frac{4}{3} ds^2(\text{S}^7/\mathbb{Z}_k) \right)\right] , ~~~e^\Phi \sim 3^{-1/4} 2^{1/4} c_0 L^3 (h''')^{-1} r^{-1/2}.
\end{equation}
This appears like the behaviour of \eqref{eq:notphysical}, however now $r\to \infty$ rather than zero. We do not recognise this behaviour as physical either, and even if it were this behaviour is at infinite proper distance from $r_0$ \textit{i.e} the interval for local solutions is semi-infinite.

The missing ingredient to construct globally bounded solutions is D8-branes on the interior of the interval, which enable us to glue  a local solution that is bounded from below to a local solution that is bounded from above at some finite value of $r$. We explore this in the next subsection.

\subsection{Global Solutions}
We now consider global solutions with D8-branes along the interior of the interval spanned by $r$.\\
~\\
As previously mentioned,  it is possible to divide the interval spanned by $r$ into distinct cells with $r\in [l,l+1]$ for which $h$ takes different values as in \eqref{hder}. When one does this it is necessary to ensure that the metric and dilaton are continuous as we cross each point $r=l$, which is ensured if we impose the same on 
\beq
h,~~~(h')^2,~~~h''.\label{eq:mustbecontinouse}
\eeq
On the other hand $h'''$ needs only be constant piecewise, with a discontinuity at $r=l$ giving rise to a $\delta$-function source as
\beq
h''''= -(\beta_{l-1}-\beta_{l})\delta(r-l).
\eeq
This has consequences for the Bianchi identities of the RR-fluxes, \textit{i.e} we find at the loci $r=l$
\beq
 dF_0= -\frac{1}{2\pi}(\beta_{l-1}-\beta_{l})\delta(r-l)dr,~~~~d\hat F_2=-\frac{1}{2}(\beta_{l-1}-\beta_{l})(r-l)\delta(r-l)dr\wedge \text{vol}(\text{AdS}_2),
\eeq
with the higher Page fluxes $F_8,F_{10}$ simply closed.  Thus, only $dF_0$ gives a non-trivial result when integrated across the  neighbourhood of $r=l$. We conclude that a discontinuity in $F_0$ at $r=l$ gives rise to a stack of $\beta_{l-1}-\beta_{l}$ D8-branes at this locus with trivial worldvolume flux.  It follows that the value $\beta_l$ in each cell should decrease as we traverse the interval towards positive infinity.\\
~\\
To construct a viable global solution that is bounded one needs to arrange for the space to begin and end at one of the physical behaviours discussed in the previous section. Between these boundaries we can place an arbitrary number of cells provided that \eqref{eq:metricposs} holds within each of them and that \eqref{eq:mustbecontinouse} are continuous as we cross between each of them. We have a stack of D8-branes at every cell intersection for which $h'''$ exhibits a discontinuity. There are in fact many ways to satisfy these conditions. Indeed \cite{Lozano:2024idt}, which studies a class of AdS$_3$ solutions also governed by an order 3 polynomial that is subject to the same constraints, explores these options in great detail. Rather than repeat that analysis here, let us explain one particularly simple way to define bounded global solutions.

Let us assume that we begin the $r$ interval at $r=0$ (\textit{i.e} at the start of the $0^{\text{th}}$ cell with $l=0$), with one of the behaviours of \eqref{eq:boundarybehaviours}. In terms of the parametrisation of \eqref{hder} this amounts to imposing one of
\begin{align}
\mathbb{R}^8/\mathbb{Z}_k&:~~~~\mu_0=\delta_0=0,~~~~~\gamma_0,\beta_0>0\nn\\[2mm]
\text{O8/D8}&:~~~~ \delta_0=\gamma_0=0,~~~~~\mu_0,\beta_0>0\nn\\[2mm]
\text{OF1}&:~~~~2\mu_0\gamma_0= \delta_0^2 ,~~~~~\mu_0,\gamma_0,\beta_0>0,\label{eq:boundarybehavioursconstants}
\end{align}
where the inequalities ensure we satisfy \eqref{eq:metricposs} for $0<r<1$. We can ensure metric continuity across an arbitrary number of cells along $\mathbb{R}^+$ by demanding that the constants appearing in \eqref{hder} obey 
\begin{align}
\mu_l&=\mu_{l-1}+\delta_{l-1}+\frac{1}{2}\gamma_{l-1}+\frac{1}{6}\beta_{l-1},\nn\\[2mm]
\delta_l&=\delta_{l-1}+\gamma_{l-1}+\frac{1}{2}\beta_{l-1},\nn\\[2mm]
\gamma_{l}&=\gamma_{l-1}+\beta_{l-1},\label{eq:transitioneqs}
\end{align}
between cells. Note that this imposes the continuity of $(h,h',h'')$, while only the continuity of $(h')^2$ is required in general.  Under these assumptions it is possible to establish that \eqref{eq:metricposs} holds in the $l^{\text{th}}$ cell if it held in the $(l-1)^{\text{th}}$ cell and $\beta_l\geq 0$. Thus the relations \eqref{eq:transitioneqs} and $\beta_l\geq 0$ guarantee a well-defined metric up to some arbitrary cell with $r\in[P-1,P]$. At the point $r=P$ one can then simply glue the profile for $h$ onto its mirror image by first imposing in the $P^{\text{th}}$ cell that
\begin{align}
\mu_P&=\mu_{P-1}+\delta_{P-1}+\frac{1}{2}\gamma_{P-1}+\frac{1}{6}\beta_{P-1},\nn\\[2mm]
\delta_P&=-(\delta_{P-1}+\gamma_{P-1}+\frac{1}{2}\beta_{P-1}),\nn\\[2mm]
\gamma_{P}&=\gamma_{P-1}+\beta_{P-1},~~~~\beta_{P}=-\beta_{P-1}\label{eq:transitioneqssignchange},
\end{align}
 where we allow $h'$ to change sign, and then for each $(P+n)^{\text{th}}$ cell for $n=1,...P-1$ after this that
\beq
\mu_{P+n}=\mu_{P-n},~~~\delta_{P+n}=-\delta_{P-n},~~~\gamma_{P+n}=\gamma_{P-n},~~~\beta_{P+n}=-\beta_{P-n-1}.
\eeq
In this way we ensure that the interval is bounded as $r\in [0,2P]$ between two physical behaviours of the same type, namely $\mathbb{R}^8/\mathbb{Z}_k$, $\text{O8/D8}$ or OF1 and that \eqref{eq:metricposs} holds globally.  Allow us to stress though that this is not the only profile that is consistent with a well defined solution, it is merely a relatively simply way to ensure that whilst maintaining the freedom of adding an arbitrary number of D8s along the interior of the interval. In general it is neither necessary for the profile of $h$ to be symmetric nor to begin and end with a common behaviour - specifically all of the well defined profiles considered (in a different context) in \cite{Lozano:2024idt} will be equally well defined here. 

\section{A new class of $\text{AdS}_2\times \mathbb{CP}^3\times \text{S}^1\times I$ solutions to Type IIB with $\mathcal{N}=6$ supersymmetry}\label{sec:N=6}
 
 In this section we present a new class of $\text{AdS}_2\times \mathbb{CP}^3\times \text{S}^1\times I$ solutions in Type IIB supergravity, constructed by acting with T-duality along the $\tau$-direction on the solutions \eqref{metric}-\eqref{RR1}, with $\tau$ as in \eqref{tau-direction}. In the next subsection we relate this class of solutions with the one constructed in \cite{Conti:2023rul}, acting with T-duality on the $\text{AdS}_3\times \mathbb{CP}^3\times I$ solutions in  \cite{Macpherson:2023cbl} along the Hopf fibre of $\text{AdS}_3$. 
 
Our new class reads 
\begin{align}
&\frac{ds^2}{L^2}=\frac{h^{3/2}\sqrt{h''}}{8(2hh''-h'^2)}ds^2(\text{AdS}_2)+\sqrt{\frac{h''}{h}}\Bigl(\frac18 dr^2+\frac{k^2}{L^4}d\tau^2\Bigr)+\sqrt{\frac{h}{h''}}ds^2(\mathbb{CP}^3),\nn\\
&e^\Phi=\frac{k\, c_0 L^2}{\sqrt{2hh''-h'^2}}\sqrt{\frac{h}{h''}}, \qquad B_2=-\frac{L^2}{8\sqrt{2}} \Bigl(\frac{hh'}{2hh''-h'^2}+r-l\Bigr)\text{vol}(\text{AdS}_2)-2k\tau J\nn,\\
&F_1=\frac{4}{c_0L^4}h''' d\tau,\label{metricIIB}\\
&\hat{F}_3=\frac{1}{c_0}\left(-\frac{1}{2\sqrt{2}L^2}\Bigl(h''-(r-l)h'''\Bigr)\text{vol}(\text{AdS}_2)+\frac{8 k}{L^4}h''' \tau J\right)\wedge d\tau,\nn \\
&\hat{F}_5=\frac{k\tau}{c_0}\left(-\frac{1}{\sqrt{2}L^2}\Bigl(h''-(r-l)h'''\Bigr)\text{vol}(\text{AdS}_2)\wedge J+ \frac{8k}{L^4}\tau h'''J\wedge J\right)\wedge d\tau,\nn
\label{RR1IIB}
\end{align}
where the gauge invariant 5-form $F_5=0$. This class of solutions preserves $\mathcal{N}=6$ supersymmetries in 1d, with OSp$(6|2)$ superconformal symmetry and SO(6) R-symmetry, that is realised geometrically on the $\mathbb{CP}^3$. These solutions provide a further new class of $\text{AdS}_2\times \mathbb{CP}^3\times \text{S}^1\times I$ solutions with non-vanishing axion field besides the ones recently constructed in  \cite{Conti:2023rul}, by acting with Abelian T-duality on the $\text{AdS}_3\times \mathbb{CP}^3\times I$ solutions in  \cite{Macpherson:2023cbl}
along the U(1) fibre of $\text{AdS}_3$. We show in the next subsection that in the limit of vanishing axion field both classes of solutions are related by S-duality.

\subsection{Massless limit. Connection to the solutions in \cite{Conti:2023rul}}\label{massless-limit}

In \cite{Dibitetto:2019nyz} it was shown that the $\text{AdS}_2\times \text{S}^7$ solution in the class in \eqref{metric}-\eqref{RR1} corresponding to vanishing Romans mass originates from the ABJM M-theory solution. In our case writing the metric of $\text{AdS}_4$ in the parametrisation
\begin{equation}\label{AdS4}
ds^2(\text{AdS}_4)=\frac{1}{\sin^2{\theta}}\Bigl(d\theta^2+ds^2(\text{AdS}_3)\Bigr)
\end{equation}
and orbifolding the  $\text{AdS}_3$ factor by $\mathbb{Z}_{k'}$ as
\begin{equation}\label{AdS3}
ds^2(\text{AdS}_3/\mathbb{Z}_{k'})=\frac14\Bigl[\Bigl(2\frac{dy}{k'}+\eta\Bigr)^2+ds^2(\text{AdS}_2)\Bigr]\qquad \text{with} \qquad d\eta=-\text{vol}(\text{AdS}_2)
\end{equation}
and $y\in [0,2\pi]$,
one arrives at the massless solution in \eqref{metric}-\eqref{RR1} with $c_0=h''/(\sqrt{2}k'L^2)$ reducing along the $y$-direction. To see this explicitly a change of variables must be performed relating the $\theta$-coordinate to the $r$-coordinate in \eqref{metric}-\eqref{RR1}. Namely, taking 
\begin{equation}\label{quadratich}
h(r)=\alpha r^2+p r+q 
\end{equation}
as the most general solution of the Bianchi identity \eqref{Bianchi} in the massless case, 
the change of variables is given by  
\begin{equation}\label{changeofcoord}
\cot{\theta}=\frac{2\alpha r+p}{\sqrt{4\alpha q-p^2}},
\end{equation}
and the ABJM M-theory metric becomes
\begin{equation}
ds^2_{\text{ABJM}}=\Bigl(\frac{(4\alpha q-p^2)k'^2L^4}{2\alpha^2}\Bigr)^{1/3}\Bigl(\frac14 ds^2(\text{AdS}_4/\mathbb{Z}_{k'})+ds^2(\text{S}^7/\mathbb{Z}_k)\Bigr).
\end{equation}

Given the previous relation, the brane intersection associated to the massless solution is the one of ABJM with extra $k'$ M-waves, propagating along the $y$-direction. 
This is depicted in Table \ref{ABJM-M0}.
\begin{table}[h]
\renewcommand{\arraystretch}{1}
\begin{center}
\scalebox{1}[1]{
\begin{tabular}{c| c cc  c c  c  c c c c c}
 branes & $x^0$ & $x^1$ & $y$ & $\tau$ & $x^3$ & $x^4$ & $x^5$ & $\psi$ & $x^7$ & $x^8$ & $x^9$ \\
\hline \hline
$\mrm{M}2$ & $\times$ & $\times$ & $\times$ & $-$ & $-$ & $-$ & $-$ & $-$ & $-$ & $-$ & $-$ \\
$\mrm{KK}'$ & $\times$ & $\times$ & $\times$ & $\times$ & $\times$ & $\times$ & $\times$ & $z$ & $-$ & $-$ & $-$ \\ 
$(\mrm{KK}',k\, \mrm{KK})$ & $\times$ & $\times$ & $\times$ & $(\times,z)$  & $\cos{\phi}$  & $\cos{\phi}$  & $\cos{\phi}$ & $(z,\times)$ & $\sin{\phi}$ & $\sin{\phi}$ & $\sin{\phi}$ \\
$k'\,\mrm{M}0$ & $\times$ & $-$ & $z$ & $-$ & $-$ & $-$ & $-$ & $-$ & $-$ & $-$ & $-$\\
\end{tabular}
}
\caption{Brane set-up associated to the $\text{AdS}_4/\mathbb{Z}_{k'}\times \text{S}^7/\mathbb{Z}_k$ solution. This is the brane intersection of ABJM with extra $k'$ units of momentum along the $y$-direction contained in $\text{AdS}_4$, in its parametrisation  \eqref{AdS4}-\eqref{AdS3}. This momentum is added including $k'$ waves propagating along the $y$-direction. The special directions of both the waves and the KK-monopoles are denoted by  $z$, to distinguish them from the rest of transverse directions.}
 \label{ABJM-M0}
\end{center}
\end{table}
As in ABJM, $\psi$ is the U(1) coordinate in the $\mathbb{CP}^3$ contained in the $\text{S}^7/\mathbb{Z}_k$ manifold in its parametrisation
\begin{equation}
ds^2(\text{S}^7/\mathbb{Z}_k)=\Bigl(\frac{d\tau}{k}+{\cal A}\Bigr)^2+ds^2(\mathbb{CP}^3),
\end{equation}
with $d{\cal A}=2J$ and $J$ the K\"ahler form of the $\mathbb{CP}^3$, and 
\begin{equation}\label{CP3}
ds^2(\mathbb{CP}^3)=d\xi^2+\frac14 \cos^2{\xi} ds^2(\text{S}^2_1)+\frac14 \sin^2{\xi}ds^2(\text{S}^2_2)+\frac14\sin^2{\xi}\cos^2{\xi}(d\psi+\eta_1+\eta_2)^2
\end{equation}
with $d\eta_i=-\text{vol}(\text{S}^2_i)$.

As it is well-known, the brane set-up that underlies the $\text{AdS}_4\times \text{S}^7/\mathbb{Z}_k$ ABJM background preserves $\mathcal{N}=3$ supersymmetries in 3d, that is, 6 Poincar\' e supercharges. Introducing the M0-branes reduces the supersymmetries by a half, giving rise to $\mathcal{N}=3$ supersymmetry in one dimension, which implies 3 Poincar\'e supercharges. As in ABJM, in the IR there should be an enhancement to $\mathcal{N}=6$ in 1d, that is, to the number of supersymmetries preserved by the solutions \eqref{metric}-\eqref{RR1}. 
Reducing along the $y$ direction the brane intersection depicted in Table \ref{ABJM-M0}  one arrives at the brane set-up that underlies the massless solution in \eqref{metric}-\eqref{RR1}, that we have depicted 
in Table \ref{ABJM-D0}. 
\begin{table}[h]
\renewcommand{\arraystretch}{1}
\begin{center}
\scalebox{1}[1]{
\begin{tabular}{c| c cc  c c  c  c c c c}
 branes & $x^0$ & $x^1$ & $\tau$ & $x^3$ & $x^4$ & $x^5$ & $\psi$ & $x^7$ & $x^8$ & $x^9$ \\
\hline \hline
$\mrm{F}1$ & $\times$ & $\times$  & $-$ & $-$ & $-$ & $-$ & $-$ & $-$ & $-$ & $-$ \\
$\mrm{KK}'$ & $\times$ & $\times$  & $\times$ & $\times$ & $\times$ & $\times$ & $z$ & $-$ & $-$ & $-$ \\ 
$(\mrm{KK}',k\,\mrm{KK})$ & $\times$  & $\times$ & $(\times,z)$  & $\cos{\phi}$  & $\cos{\phi}$  & $\cos{\phi}$ & $(z,\times)$ & $\sin{\phi}$ & $\sin{\phi}$ & $\sin{\phi}$ \\
$k'\,\mrm{D}0$ & $\times$ & $-$  & $-$ & $-$ & $-$ & $-$ & $-$ & $-$ & $-$ & $-$\\
\end{tabular}
}
\caption{Brane intersection associated to the $\text{AdS}_2\times I \times \text{S}^7/\mathbb{Z}_k$ massless solution in the class \eqref{metric}-\eqref{RR1}. $z$ is the Taub-NUT direction of the KK-monopoles. This brane intersection is obtained reducing along the $y$-direction the intersection depicted in Table \ref{ABJM-M0}.}
 \label{ABJM-D0}
\end{center}
\end{table}
T-dualising now along the $\tau$-direction one finds the brane set-up depicted in Table \ref{ABJM-D1}. The dual background is a new $\text{AdS}_2\times I \times \text{S}^1/\mathbb{Z}_k\times \mathbb{CP}^3$ solution to Type IIB supergravity that is contained in the class of \eqref{metricIIB}. 
\begin{table}[h]
\renewcommand{\arraystretch}{1}
\begin{center}
\scalebox{1}[1]{
\begin{tabular}{c| c cc  c c  c  c c c c}
 branes & $x^0$ & $x^1$ & $\tau$ & $x^3$ & $x^4$ & $x^5$ & $\psi$ & $x^7$ & $x^8$ & $x^9$ \\
\hline \hline
$\mrm{F}1$ & $\times$ & $\times$  & $-$ & $-$ & $-$ & $-$ & $-$ & $-$ & $-$ & $-$ \\
$\mrm{KK}'$ & $\times$ & $\times$  & $\times$ & $\times$ & $\times$ & $\times$ & $z$ & $-$ & $-$ & $-$ \\ 
$(\mrm{KK}',k\,\mrm{NS}5)$ & $\times$  & $\times$ & $(\times,-)$  & $\cos{\phi}$  & $\cos{\phi}$  & $\cos{\phi}$ & $(z,\times)$ & $\sin{\phi}$ & $\sin{\phi}$ & $\sin{\phi}$ \\
$k'\,\mrm{D}1$ & $\times$ & $-$  & $\times$ & $-$ & $-$ & $-$ & $-$ & $-$ & $-$ & $-$\\
\end{tabular}
}
\caption{Brane intersection associated to the $\text{AdS}_2\times I\times \text{S}^1/\mathbb{Z}_k\times \mathbb{CP}^3$ solution in the class of \eqref{metricIIB} for vanishing axion field. $z$ indicates the Taub-NUT direction of the KK-monopoles. This brane intersection is obtained T-dualising along the $\tau$-direction the brane intersection depicted in Table \ref{ABJM-D0}.}
 \label{ABJM-D1}
\end{center}
\end{table}

We can now see that this solution is related by S-duality to the solution corresponding to vanishing axion field in the general class of $\text{AdS}_2\times \mathbb{CP}^3\times \text{S}^1\times I$ Type IIB backgrounds recently constructed in  \cite{Conti:2023rul}. This general class was derived by acting with Abelian T-duality on the $\text{AdS}_3\times \mathbb{CP}^3\times I$ solutions in massive Type IIA supergravity presented in \cite{Macpherson:2023cbl}, with the T-duality performed along the S$^1$ fibre of  $\text{AdS}_3$, in its parametrisation \eqref{AdS3}. The solution with vanishing axion field is related by T-duality to the $\text{AdS}_3\times \mathbb{CP}^3\times I$ solution in \cite{Macpherson:2023cbl} corresponding to vanishing Romans mass, which in turn was  shown to be related to the $\text{AdS}_4\times \mathbb{CP}^3$ solution of ABJM/ABJ  through a change of variables \cite{Lozano:2024idt}. 

More concretely, the $\text{AdS}_3\times \mathbb{CP}^3\times I$ solutions in \cite{Macpherson:2023cbl} are defined in terms of the same $h(r)$ function satisfying \eqref{Bianchi}. For vanishing Romans mass this is a quadratic polynomial that can be written as in \eqref{quadratich}, with its second derivative giving the D6-brane charge, namely $2\alpha=k$.
Taking this $h$, the same change of coordinates given by \eqref{changeofcoord}
gives rise to $\text{AdS}_4\times \mathbb{CP}^3$, with $\text{AdS}_4$ parametrised as in \eqref{AdS4}. In this section we take 
\begin{equation}
L^2=\frac{4\pi}{\sqrt{2}}\frac{k}{k'}
\end{equation}
to match the conventions in  \cite{Conti:2023rul}.
After the T-duality along the Hopf-fibre of $\text{AdS}_3$, that we previously mod out by $\mathbb{Z}_{k'}$,  and rescaling $g_s$ such that $(e^{-\Phi},F_-)\to  \frac{2\alpha}{k}(e^{-\Phi},F_-)$ we obtain
\begin{align}
&ds^2=\frac{\pi  \sqrt{4\alpha q-p^2}}{8\alpha\sin^2{\theta}}\Bigl[ds^2(\text{AdS}_2)+16\sin^2{\theta}\,ds^2(\mathbb{CP}^3)+4d\theta^2+\frac{16\alpha^2 k'^2}{\pi^2(4\alpha q-p^2)}\sin^4{\theta}\,dy^2\Bigr]\nn\\[2mm]
&e^{\Phi}=\frac{2k'\sin{\theta}}{k}, \qquad H_3=-\frac{k'}{2} dy\wedge \text{vol}(\text{AdS}_2)\nn \\
&F_3= -\frac{3\pi k\sqrt{4\alpha q-p^2}}{16k'\alpha\sin^4{\theta}}d\theta\wedge\text{vol}(\text{AdS}_2)+2k\, dy\wedge J\, , \qquad F_5=0. \label{IIBmassless1}
\end{align}
One can now simply check that this solution, that lies in the class in \cite{Conti:2023rul} for vanishing axion field,  is S-dual to the solution in  \eqref{metricIIB}, with $h$ given by \eqref{quadratich} and $r$ mapped onto $\theta$ via the change of coordinates \eqref{changeofcoord}. This solution is given by
\begin{align}
&ds^2=\frac{\pi k\sqrt{4\alpha q-p^2}}{16\alpha k'\sin^3{\theta}}\Bigl[ds^2(\text{AdS}_2)+16\sin^2{\theta}\,ds^2(\mathbb{CP}^3)+4d\theta^2+\frac{16\alpha^2 k'^2}{\pi^2(4\alpha q-p^2)}\sin^4{\theta}\,d\tau^2\Bigr]\nn\\
&e^{\Phi}=\frac{k}{2k'\sin{\theta}}, \qquad H_3=\frac{3\pi k\sqrt{4\alpha q-p^2}}{16k'\alpha\sin^4{\theta}}d\theta\wedge\text{vol}(\text{AdS}_2)-2k\, d\tau\wedge J \nn\\
&F_3=-\frac{k'}{2} d\tau\wedge \text{vol}(\text{AdS}_2)\, , \qquad F_5=0,\label{IIBmassless2}
\end{align}
and indeed one can easily confirm that, upon fixing $y=\tau$, the solution of \eqref{IIBmassless1} is mapped into that of \eqref{IIBmassless2} via
\beq
\Phi\to-\Phi,~~~~ ds^2\to e^{-\Phi}ds^2,~~~~ \left(\begin{array}{c}H_3\\F_3\end{array}\right)\to \left(\begin{array}{cc}0&-1\\1&0\end{array}\right)\left(\begin{array}{c}H_3\\F_3\end{array}\right),
\eeq
which are precisely the string frame S-duality transformation rules for solutions with trivial axion field. We already noted that before the T-duality $\alpha$ in \eqref{IIBmassless1} is related to $k$, the charge of D6-branes. This suggests that we should fix
\beq
2\alpha=k,\label{eq:fixalpha}
\eeq
in both \eqref{IIBmassless1} and \eqref{IIBmassless2}. As a consistency check one can see that the brane intersections that underlie both solutions are also related by S-duality. The brane set-up associated to the solution \eqref{IIBmassless1} is the one depicted in Table  \ref{Andreasol}. It is obtained T-dualising along the $y$-direction the brane set-up underlying the $\text{AdS}_3$ solutions in \cite{Macpherson:2023cbl}, proposed in \cite{Lozano:2024idt}, particularised to vanishing Romans mass.
\begin{table}[h]
\renewcommand{\arraystretch}{1}
\begin{center}
\scalebox{1}[1]{
\begin{tabular}{c| c cc  c c  c  c c c c}
 branes & $x^0$ & $y$ & $r$ & $x^3$ & $x^4$ & $x^5$ & $\psi$ & $x^7$ & $x^8$ & $x^9$ \\
\hline \hline
$\mrm{D}1$ & $\times$ & $-$  & $\times$ & $-$ & $-$ & $-$ & $-$ & $-$ & $-$ & $-$ \\
$\mrm{KK}'$ & $\times$ & $\times$  & $\times$ & $\times$ & $\times$ & $\times$ & $z$ & $-$ & $-$ & $-$ \\ 
$(\mrm{KK}',k\,\mrm{D5})$ & $\times$  & $(\times,-)$ & $\times$  & $\cos{\phi}$  & $\cos{\phi}$  & $\cos{\phi}$ & $(z,\times)$ & $\sin{\phi}$ & $\sin{\phi}$ & $\sin{\phi}$ \\
$k'\,\mrm{F}1$ & $\times$ & $\times$  & $-$ & $-$ & $-$ & $-$ & $-$ & $-$ & $-$ & $-$\\
\end{tabular}
}
\caption{Brane intersection associated to the $\text{AdS}_2\times \mathbb{CP}^3\times \text{S}^1/\mathbb{Z}_{k'}\times I$ solution \eqref{IIBmassless1}. $y$ is the direction along the $\text{S}^1/\mathbb{Z}_{k'}$, $r$ parametrises the interval, $\psi$ is the direction of the $\mathbb{CP}^3$ in its parametrisation \eqref{CP3}, and $(x^3,x^4,x^5)$ and $(x^7,x^8,x^9)$ are the directions realising its $SO(3)\times SO(3)$ isometry group.}
 \label{Andreasol}
\end{center}
\end{table}
 In turn, the brane intersection  underlying the solution \eqref{IIBmassless2} is the one depicted in Table \ref{ABJM-D1}. One can check that both brane set-ups are related by S-duality. 
 
The relation between the two solutions under S-duality clarifies as well the role played by the constants $p$ and $q$, introduced in equation \eqref{quadratich}, in the solution  of \eqref{IIBmassless2}. They are integers associated to the numbers of additional, fractional, D1-branes and the F1-strings present in the configuration. The solution of \eqref{IIBmassless1} allows for a NS 2-form potential of the form
 \begin{equation}
 B_2=-\frac{1}{2}k'(y-l)\text{vol}(\text{AdS}_2)+(B_2)_{\text{flat}},~~~~ (B_2)_{\text{flat}}=4\pi\,\frac{p}{k}J,\label{eq:b2forads3tdual}
 \end{equation}
where $(B_2)_{\text{flat}}$ can be obtained via T-duality from the flat $B_2$-field of the ABJ theory. This allows us to define Page fluxes via $\hat{F}=e^{-B_2}\wedge F$ for the RR 5 and 7-forms. Focusing on their magnetic parts we find
\beq
\hat F_5^{\text{mag}}=-8\pi p\, dy\wedge J\wedge J,~~~~\hat F_7^{\text{mag}}=32\pi^2 q\, dy\wedge J\wedge J\wedge J.
\eeq
Upon S-duality \eqref{eq:b2forads3tdual} gives rise to the  RR 2-form potential of $F_3$ in \eqref{IIBmassless2},
 \begin{equation}
 C_2=-\frac{1}{2}k'(\tau-l)\text{vol}(\text{AdS}_2)+(C_2)_{\text{flat}},~~~~(C_2)_{\text{flat}}=4\pi\, \frac{p}{k}J.
 \end{equation}
In turn, this gives rise to the Page fluxes for $F_5$ and $H_7=e^{-2\Phi}\star H_3$ in \eqref{IIBmassless2} defined as
 \begin{equation}
 \hat{F}_5=F_5+H_3\wedge C_2,~~~~\hat{H}_7=H_7+F_3\wedge C_4+\frac{1}{2}H_3\wedge C_2\wedge C_2.
 \end{equation}
Focusing again on only the magnetic components  we find
\beq
\hat F_5^{\text{mag}}=-8\pi p\, d\tau\wedge J\wedge J,~~~~\hat H_7^{\text{mag}}=32\pi^2 q\, d\tau\wedge J\wedge J\wedge J.
\eeq
Therefore for both the solutions of \eqref{IIBmassless1} and \eqref{IIBmassless2} we can define a D3-brane charge
 \begin{equation}
 Q_{{\rm D}3}=\frac{1}{(2\pi)^4}\int \hat{F}_5=p.
 \end{equation}
This represents an additional fractional D1-brane charge present in the configuration, that plays the same role as the 
 fractional D3-brane charge of the ABJM theory. In turn, we have from \eqref{IIBmassless2},
 \begin{equation}
 Q_{{\rm F}1}=\frac{1}{(2\pi)^6}\int \hat H_7=q.
 \end{equation}
 Therefore $q$ gives the total number of F1-strings in the brane set-ups described by Tables \ref{ABJM-D0} and \ref{ABJM-D1}.

Figure \ref{fig:dualities} summarises the relation between the two solutions discussed in this subsection.

\section{Field theory interpretation of the $\mathcal{N}=8$ solutions}\label{sec:N=8FT}

The brane set-up in Table \ref{F1-D0-D8} is our starting point for the field theory interpretation of the $\mathcal{N}=8$ solutions. In this section we show that the D0-branes are interpreted, together with the F1-strings, as baryon vertices for the D8-branes. 

\begin{figure}[h]
\centering
\includegraphics[scale=0.65]{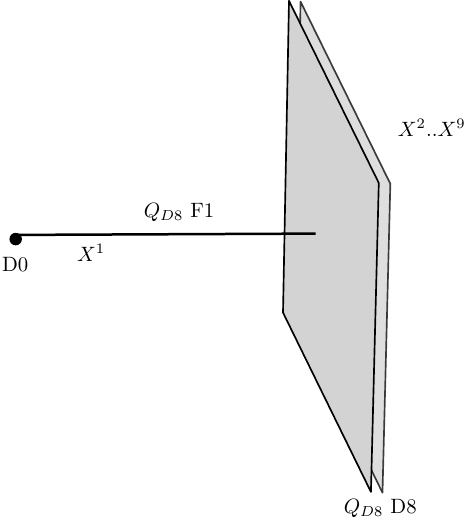}
\caption{F1-strings stretched between D0 and D8-branes.}\label{Wilson-lines}
\end{figure}

The worldvolume coupling in the WZ action of a D0-brane,
\begin{equation}
S_{{\rm D}0}=T_0\int F_0\wedge A_t=T_{F1}Q_{{\rm D}8}\int A_t
\end{equation}
shows that in the presence of a Romans mass the D0-brane captures $Q_{{\rm D}8}$ units of F1-string charge, and therefore behaves as a baryon vertex for this number of D8-branes. Indeed, the relative orientation between the D0 and the D8-branes in the brane set-up in Table \ref{F1-D0-D8} is the one that allows to create F1-strings stretched between them, as depicted in Figure \ref{Wilson-lines}.
These strings have as their lowest energy excitation a fermionic field, which upon integration leads to a Wilson loop  \cite{Rey:1998ik,Maldacena:1998im}.
As shown in \cite{Yamaguchi:2006tq,Gomis:2006sb}, in order to describe a half-BPS Wilson loop in an antisymmetric representation labelled by a Young tableau with columns $(n_1,n_2,\dots,n_M)$, a configuration of stacks of branes separated a distance $L$ from the colour branes must be considered, with $(n_1,n_2,\dots,n_M)$ F1-strings stretched between each stack and the colour branes. This generalises the description of a Wilson loop in \cite{Rey:1998ik,Maldacena:1998im} to arbitrary antisymmetric representations. 

\begin{figure}[h]
\centering
\includegraphics[scale=0.8]{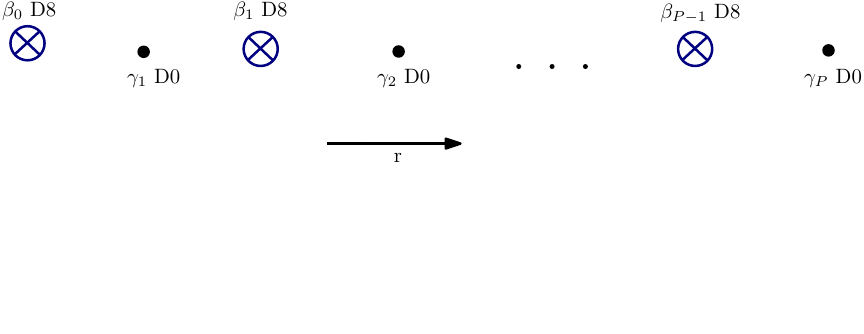}
\vspace{-2cm}
\caption{Hanany-Witten like brane set-up associated to the quantised charges of the $\mathcal{N}=8$ solutions.}\label{D0-F1-D8branes}
\end{figure}

\begin{figure}[h]
\centering
\includegraphics[scale=0.85]{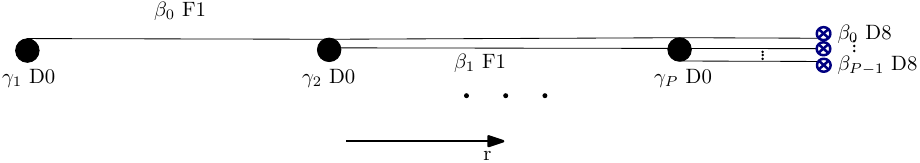}
\caption{Hanany-Witten like brane set-up equivalent to the configuration in Figure \ref{D0-F1-D8branes}.}\label{D0-F1-D8branesfinal}
\end{figure}

In the particular configuration associated to the $\mathcal{N}=8$ solutions there are $Q^l_{{\rm D}0}=\gamma_l$ D0 and $Q^l_{{\rm D}8}=\beta_l$ D8-branes in the $r\in [l,l+1]$ interval, as depicted in Figure \ref{D0-F1-D8branes}.  Here we assume that the space starts at $r=0$ and ends at $r=P$, with any of the physically meaningful possibilities discussed in section 2.2. One such possibility is to glue the quiver to itself, as explained in that section. 
Following the same analysis in \cite{Lozano:2020sae} this  can be related
by a combination of a T-duality, an S-duality, successive Hanany-Witten moves and a further T-duality to the brane configuration depicted in Figure \ref{D0-F1-D8branesfinal}, from where the relation with the construction in  \cite{Yamaguchi:2006tq,Gomis:2006sb} becomes manifest (see \cite{Lozano:2020sae} for the details of this analysis). 
Figure \ref{D0-F1-D8branesfinal} shows that the sum of the F1-strings stretched between each stack of D0-branes and the stacks of D8 flavour branes  coincides with the rank of the gauge group living in the D0-branes, given the condition \eqref{betagamma}. This implies that the Wilson lines are actually in the fundamental representation of the gauge groups. Therefore, the D0-branes describe baryon vertices for the D8-branes.

The conclusion of this analysis is that the $\text{AdS}_2\times \text{S}^7\times I$ solutions can be interpreted as describing backreacted baryon vertices for the 9d field theories living in the D8-branes. As in similar examples discussed in the literature (see \cite{Lozano:2020sae,Ramirez:2021tkd,Lozano:2021fkk,Lozano:2022vsv}), adding the vertices turns the gauge groups living in the higher dimensional branes into flavour groups, with the branes interpreted as vertices becoming the colour branes of the new backreacted geometry. If the resulting theory is conformal the backreacted geometry is of $\text{AdS}_2$-type close to the horizon. In the examples  discussed in \cite{Lozano:2020sae,Ramirez:2021tkd,Lozano:2021fkk,Lozano:2022vsv} the higher dimensional theory in which the vertices were inserted was already conformal, and dual to a higher dimensional AdS space. A new feature of our construction in this section is that the gauge theory living in the D0-F1-D8-branes flows to a conformal quantum mechanics (in the IR) even if the theory living in the D8-branes is not conformal.

\subsection{Quiver quantum mechanics and central charge}

$\mathcal{N}=8$ superconformal quantum mechanics with OSp$(8|2)$ supergroup have been less explored in the literature than their  counterparts with SU$(1,1|4)$, OSp$(4|4)$ or F(4) supergroups. Only recently there has been significant progress in the construction of Hamiltonians and the development of a unified treatment suitable for all supergroups (see \cite{Fedoruk:2024osh,Khastyan:2024fqh,Krivonos:2024onh}). Yet, the number of non-trivial and physically relevant examples described by these theories remains very scarce. In \cite{Okazaki:2015pfa} it was shown that an $\mathcal{N}=16$ superconformal quantum mechanics with OSp$(16|2)$ supergroup arises upon wrapping the BLG model on a 2-torus. Similarly, we have shown that an $\mathcal{N}=8$ superconformal quantum mechanics with OSp$(8|2)$ supergroup arises upon adding momentum to the M2-branes that underlie the BLG model, which implies a reduction of the number of supersymmetries by a half. Once this configuration is reduced to Type IIA, extra D8-branes can be added without any further breaking of the supersymmetries and global symmetries.
Remarkably, the $\text{AdS}_2\times \text{S}^7\times I$ solutions constructed in \cite{Dibitetto:2018gbk} provide the explicit geometric backgrounds dual to these SCQMs, which should then play a prominent role in the microscopical description of black holes with these near-horizon geometries.

\begin{figure}[h]
\centering
\includegraphics[scale=0.6]{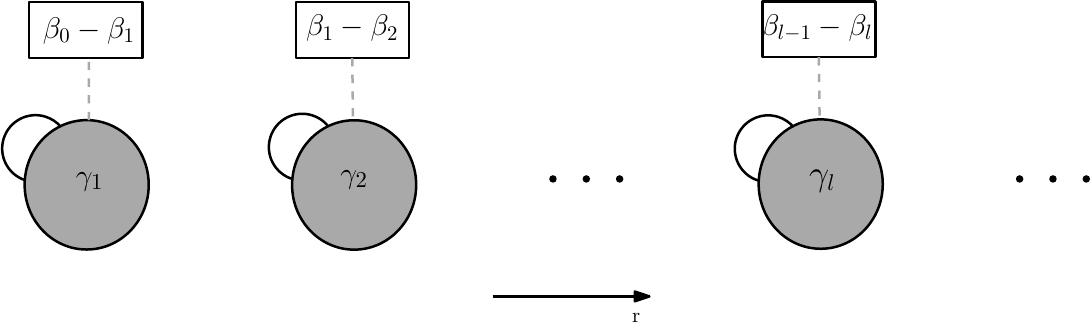}
\caption{Disconnected quivers describing the superconformal quantum mechanics dual to the solutions \eqref{metric}-\eqref{RR1} for $k=1$. Circles denote $\mathcal{N}=8$ vector multiplets, black lines $\mathcal{N}=8$ hypermultiplets and grey dashed lines $\mathcal{N}=2$ Fermi multiplets.}\label{D0-F1-D8quiver}
\end{figure}

Our proposed quantum mechanics can be given a quiver-like description in terms of a set of disconnected quivers, depicted in Figure \ref{D0-F1-D8quiver}, showing the interactions between the D0 and the D8-branes in each $r$-interval. 
In these quivers circles denote $\mathcal{N}=8$ vector multiplets, black lines adjoint $\mathcal{N}=8$ hypermultiplets and dashed lines fundamental $\mathcal{N}=2$ Fermi multiplets.  In principle it should be possible to compute the central charge associated to these quivers and check whether it matches the corresponding holographic expression. 

The definition of central charge in superconformal quantum mechanics is however subtle. Namely, in a 1d theory there is only one component of $T_{\mu\nu}$, so if the theory is conformal the vanishing of the trace implies that $T_{tt}=0$. Different proposals for counting the degrees of freedom of 1d SCFTs have been proposed in the literature (see \cite{Hartman:2008dq,Alishahiha:2008tv,Cadoni:1999ja,Balasubramanian:2009bg,vanBeest:2020vlv,Denef:2002ru,Cordova:2014oxa,Ohta:2014ria}). In particular, when the superconformal quantum mechanics is obtained by dimensional reduction of a 2d chiral CFT one can define the central charge as the central extension of the superconformal algebra, by analogy with the 2d case. This is the definition that is of interest to us in this paper, since $\mathcal{N}=4$ multiplets can be obtained by dimensional reduction of 2d $\mathcal{N}=(0,4)$ multiplets. The SCQM central charge can then be computed as the central extension of the  OSp$(8|2)$ superconformal algebra, given by 
\cite{Bershadsky:1986ms}
\begin{equation}\label{Bersha}
c=c_R=3k\,\Bigl(\frac{k+9}{k+5}\Bigr),
\end{equation}
where $k$ is the level of the 2d algebra. In turn, for superconformal algebras with $(0,2)$ supersymmetry the level of the algebra can be computed from the U(1)$_R$ R-symmetry anomaly, using that
\begin{equation}\label{level}
k=\text{Tr}[\gamma_3 Q_R^2],
\end{equation}
where $Q_R$ is the R-charge under the U(1)$_R$ R-symmetry group, and the trace is taken over all Weyl fermions in the theory. 
In 2d language the R-charges of the different fields that couple in the quiver depicted in Figure \ref{D0-F1-D8quiver} are summarised in Table \ref{Table:R-charges}, for the decomposition of $\mathcal{N}=8$ multiplets onto $\mathcal{N}=4$ multiplets\footnote{A $\mathcal{N}=8$ vector multiplet is decomposed into a $\mathcal{N}=4$ vector multiplet and two $\mathcal{N}=2$ (adjoint) Fermi multiplets, and a $\mathcal{N}=8$ hypermultiplet is decomposed into three $\mathcal{N}=4$ hypermultiplets. This will be further discussed in subsection \ref{sec:N=6FT}.}. 
\begin{table}[h]
\begin{center}
\begin{tabular}{|c|c|c|}
\hline
Multiplet & chirality & R-charge\\
\hline\hline
(0,4) hyper &  R.H. & $-1$ \\
\hline
(0,4) twisted hyper & R.H. & $0$ \\
\hline
(0,4) vector  & L.H. & $1$ \\
\hline
(0,2) Fermi& L.H. & $0$ \\
\hline
\end{tabular}
\end{center}
\caption{R-charges and chiralities of fermions in (0,4) multiplets.}
\label{Table:R-charges}
\end{table}

In turn, the holographic central charge can be computed from the solutions \eqref{metric}-\eqref{NSNS1} using the Brown-Henneaux formula
\begin{equation}
c_{hol}=\frac{3V_{int}}{4\pi G_N},
\end{equation}
where $G_N$ is the ten dimensional Newton's constant, $G_N=8\pi^6$, and $V_{int}$ is the volume of the internal manifold, which in the presence of a dilaton-field is computed from \cite{Klebanov:2007ws,Macpherson:2014eza}
\begin{equation}
V_{int}=\int d^8x\,e^{-2\Phi}\sqrt{\text{det}g_8}.
\end{equation}
This gives
\begin{equation}
c_{hol}=\frac{2^7}{\pi^3}\text{Vol}(\text{S}^7)\int dr (2hh''-h'^2)\label{chol}
\end{equation}
and substituting $h_l$ from \eqref{hder},
\begin{equation}
c_{hol}=\frac{2^7}{\pi^3}\text{Vol}(\text{S}^7)\sum_{l=0}^P\Bigl(2\mu_l\gamma_l-\delta_l^2-\beta_l(\mu_l+\frac13 \delta_l+\frac{1}{12}\gamma_l-\frac{1}{60})\Bigr).
\end{equation}
This quantity should match the calculation of $c$ from \eqref{Bersha}-\eqref{level}, in the limit in which the $\text{AdS}_2$ solutions \eqref{metric}-\eqref{NSNS1} are trustable, that is, when the ranks of the gauge groups in the quiver are large and so is the number of nodes, such that the flavour branes are sparse enough that they don't backreact in the geometry. However, in contrast with the expression for $c_{hol}$, one can see that the expression for $k$ in \eqref{level} depends only on $\gamma_l$, given that the fermions in the Fermi multiplets have vanishing R-charge. Even if one could expect that a mixing with global symmetries could occur that would change the IR R-current,  there are no Abelian global symmetries in the theory, so this does not seem like a possible way out to match both expressions. Reproducing \eqref{chol} from the field theory side remains an open problem that would be interesting to further investigate.

\section{Field theory interpretation of the $\mathcal{N}=6$ solutions}\label{sec:N=6FT}

$\mathcal{N}=6$ superconformal quantum mechanics are exotic quantum mechanics that can occur with two supergroups, SU$(1,1|3)$ and OSp$(6|2)$. Neither of these realisations seems to have been explored in much detail in the literature.
In \cite{Okazaki:2015pfa} an $\mathcal{N}=12$ superconformal quantum mechanics with SU$(1,1|6)$ supergroup was shown to arise  upon wrapping the ABJM model on a 2-torus.
Similarly, our findings show that an $\mathcal{N}=6$ superconformal quantum mechanics with OSp$(6|2)$ supergroup should arise upon adding momentum to the M2-branes probing a $\mathbb{C}^4/\mathbb{Z}_k$ singularity that underlie the ABJM model. This implies a reduction of the number of supersymmetries by a half. 
Once this is reduced to Type IIA extra D8-branes can be added without any further breaking of the supersymmetries and global symmetries.
Remarkably, the Type IIA $\text{AdS}_2\times \text{S}^7/\mathbb{Z}_k\times I$ and Type IIB $\text{AdS}_2\times \mathbb{CP}^3\times \Sigma_2$ solutions constructed in section \ref{sec:N=6} \footnote{Also the solutions in  \cite{Conti:2023rul}.} provide explicit geometric realisations of these SCQMs. Therefore, it is expected that they should play a prominent role in the microscopical description of black holes with these near-horizon geometries.

In section \ref{sec:N=6} a relation has been found between the $\text{AdS}_2\times \text{S}^7/\mathbb{Z}_k\times I$ solutions with vanishing Romans mass (and the $\text{AdS}_2\times \mathbb{CP}^3\times \Sigma_2$ solutions related to them by T-duality) and the ABJM theory. Inspired by this connection, we will make the assumption that the $\mathcal{N}=6$ superconformal quantum mechanics dual to the solutions for non-vanishing Romans mass arises as well in the IR limit of the $\mathcal{N}=3$ supersymmetric quantum mechanics living in the underlying brane intersection. We will construct this brane set-up in Type IIA, after some T-dualities are performed that map the Kaluza--Klein monopoles onto NS5-branes. At the level of the solutions these T-dualities reduce the supersymmetry from $\mathcal{N}=6$ to $\mathcal{N}=4$, exactly as it happens in the ABJM theory, but with half the number of supersymmetries. One further assumption that we will make is that, as in ABJM, the $\mathcal{N}=3$ supersymmetric quantum mechanics can be studied from the field content of the $\mathcal{N}=4$ supersymmetric quantum mechanics that arises before the rotations of some of the branes in the brane set-up are considered. Indeed, as in 3d, $\mathcal{N}=4$ supersymmetric quantum mechanics have been much more explored in the literature than their $\mathcal{N}=3$ counterparts. Of particular relevance to our constructions in this paper are the realisations of $\mathcal{N}=4$ quantum mechanics as dimensional reductions of 2d $(0,4)$ gauge theories, and the emergence of the latter in brane-box constructions \cite{Hanany:2018hlz,Faedo:2020lyw,Lozano:2024idt}.

We start discussing the brane set-up associated to the $\text{AdS}_2\times \text{S}^7/\mathbb{Z}_k\times I$ solutions of massive IIA and the new class of $\text{AdS}_2\times \mathbb{CP}^3\times \text{S}^1/\mathbb{Z}_k\times I$ solutions given by  \eqref{metricIIB}, related to them by T-duality. We consider the general case $h'''\neq 0$. In order to do so we use the Type IIA description that is obtained by T-dualising the Type IIB brane set-up along the $\psi$-direction. In the massless case this is given by the brane intersection depicted in Table \ref{ABJM-D2}. 
\begin{table}[h]
\renewcommand{\arraystretch}{1}
\begin{center}
\scalebox{1}[1]{
\begin{tabular}{c| c cc  c c  c  c c c c}
 branes & $x^0$ & $x^1$ & $\tau$ & $x^3$ & $x^4$ & $x^5$ & $\psi$ & $x^7$ & $x^8$ & $x^9$ \\
\hline \hline
$\mrm{F}1$ & $\times$ & $\times$  & $-$ & $-$ & $-$ & $-$ & $-$ & $-$ & $-$ & $-$ \\
$\mrm{NS}5'$ & $\times$ & $\times$  & $\times$ & $\times$ & $\times$ & $\times$ & $-$ & $-$ & $-$ & $-$ \\ 
$(\mrm{NS}5',k\,\mrm{NS}5)$ & $\times$  & $\times$ & $(\times,-)$  & $\cos{\phi}$  & $\cos{\phi}$  & $\cos{\phi}$ & $(-,\times)$ & $\sin{\phi}$ & $\sin{\phi}$ & $\sin{\phi}$ \\
$k'\,\mrm{D}2$ & $\times$ & $-$  & $\times$ & $-$ & $-$ & $-$ & $\times$ & $-$ & $-$ & $-$\\
\end{tabular}
}
\caption{Brane intersection associated to the $\text{AdS}_2\times \mathbb{CP}^3 \times \text{S}^1/\mathbb{Z}_k\times I$ solution  \eqref{IIBmassless2}, in its ``natural'' Type IIA realisation, where the KK-monopoles have been transformed into NS5-branes.}
 \label{ABJM-D2}
\end{center}
\end{table}
This configuration preserves $\mathcal{N}=4$ supersymmetries in 1d, previous to the rotation of the (NS5$'$,NS5) bound states with respect to the NS5$'$-branes located at $\psi=0,2\pi$. After the rotation supersymmetry is reduced to $\mathcal{N}=3$, which should be enhanced to $\mathcal{N}=4$ in the IR, the number of supersymmetries preserved by the T-dual along $\psi$ of the  $\mathcal{N}=6$
$\text{AdS}_2\times \mathbb{CP}^3\times \text{S}^1/\mathbb{Z}_k\times I$ solution. 

In the general case $h'''\neq 0$ there are extra D8-branes in the $\text{AdS}_2\times \text{S}^7/\mathbb{Z}_k\times I$ solutions and extra D7-branes in the $\text{AdS}_2\times \mathbb{CP}^3\times \text{S}^1/\mathbb{Z}_k\times I$ solutions. The brane intersection depicted in Table \ref{ABJM-D0} becomes then the one depicted in Table \ref{tablen}. In this set-up the D8-branes do not break any further supersymmetries.
\begin{table}[h]
\renewcommand{\arraystretch}{1}
\begin{center}
\scalebox{1}[1]{
\begin{tabular}{c| c cc  c c  c  c c c c}
 branes & $x^0$ & $x^1$ & $\tau$ & $x^3$ & $x^4$ & $x^5$ & $\psi$ & $x^7$ & $x^8$ & $x^9$ \\
\hline \hline
$\mrm{F}1$ & $\times$ & $\times$  & $-$ & $-$ & $-$ & $-$ & $-$ & $-$ & $-$ & $-$ \\
$\mrm{KK}'$ & $\times$ & $\times$  & $\times$ & $\times$ & $\times$ & $\times$ & $z$ & $-$ & $-$ & $-$ \\ 
$(\mrm{KK}',k\,\mrm{KK})$ & $\times$  & $\times$ & $(\times,z)$  & $\cos{\phi}$  & $\cos{\phi}$  & $\cos{\phi}$ & $(z,\times)$ & $\sin{\phi}$ & $\sin{\phi}$ & $\sin{\phi}$ \\
$\mrm{D}0$ & $\times$ & $-$  & $-$ & $-$ & $-$ & $-$ & $-$ & $-$ & $-$ & $-$\\
$\mrm{D}8$ & $\times$ & $-$ & $\times$ & $\times$ & $\times$ & $\times$ & $\times$ & $\times$ & $\times$ & $\times$ \\
\end{tabular}
}
\caption{Brane intersection associated to the $\text{AdS}_2\times \text{S}^7/\mathbb{Z}_k\times I $ solutions for non-zero Romans mass.}
 \label{tablen}
\end{center}
\end{table}
T-dualising along the $\tau$-direction contained in $\text{S}^7/\mathbb{Z}_k$ and, further, along the $\psi$ direction contained in the $\mathbb{CP}^3$, the ``natural'' brane set-up in which the Kaluza--Klein monopoles have been mapped onto NS5-branes, is obtained. We have depicted it in Table \ref{tablendualised}.
\begin{table}[h]
\renewcommand{\arraystretch}{1}
\begin{center}
\scalebox{1}[1]{
\begin{tabular}{c| c cc  c c  c  c c c c}
 branes & $x^0$ & $x^1$ & $\tau$ & $x^3$ & $x^4$ & $x^5$ & $\psi$ & $x^7$ & $x^8$ & $x^9$ \\
\hline \hline
$\mrm{F}1$ & $\times$ & $\times$  & $-$ & $-$ & $-$ & $-$ & $-$ & $-$ & $-$ & $-$ \\
$\mrm{NS}5'$ & $\times$ & $\times$  & $\times$ & $\times$ & $\times$ & $\times$ & $-$ & $-$ & $-$ & $-$ \\ 
$(\mrm{NS}5',k\,\mrm{NS}5)$ & $\times$  & $\times$ & $(\times,-)$  & $\cos{\phi}$  & $\cos{\phi}$  & $\cos{\phi}$ & $(-,\times)$ & $\sin{\phi}$ & $\sin{\phi}$ & $\sin{\phi}$ \\
$\mrm{D}2$ & $\times$ & $-$  & $\times$ & $-$ & $-$ & $-$ & $\times$ & $-$ & $-$ & $-$\\
$\mrm{D}6$ & $\times$ & $-$ & $-$ & $\times$ & $\times$ & $\times$ & $-$ & $\times$ & $\times$ & $\times$ \\
\end{tabular}
}
\caption{``Natural'' brane set-up associated to the $\text{AdS}_2\times \text{S}^7/\mathbb{Z}_k\times I $ solutions for non-zero Romans mass, with the Kaluza--Klein monopoles mapped onto NS5-branes by T-duality.}
 \label{tablendualised}
\end{center}
\end{table}
This brane set-up describes a brane box model \cite{Hanany:1997tb,Hanany:1998it,Hanany:2018hlz}, in which D2 colour branes stretch between orthogonal NS5 and NS5$'$-branes located along the $\tau$ and $\psi$ directions. On top of this there are orthogonal, flavour, D6-branes at each $\tau$, $\psi$ interval. 

It is noteworthy that already in the massless case, that is, when $h'''=0$ and the brane set-up reduces to the one depicted in Table \ref{ABJM-D2}, the field theory lives in a D2-NS5-NS5$'$ brane box. This  can then be used to find the quantum mechanics that describes M2-branes probing a $\mathbb{C}^4/\mathbb{Z}_k$ singularity with momentum, which we now turn to describe.

\subsection{M2-branes on $\mathbb{C}^4/\mathbb{Z}_k$ with momentum as a brane box}

An interesting output of our study is that it is possible to add momentum to the ABJM model in such a way that a superconformal quantum mechanics with $\mathcal{N}=6$ supersymmetry remains, and that this SCQM arises as the IR fixed point of a field theory living in a D2-NS5-NS5$'$ brane box. In this subsection we elaborate on this description.

Brane box models have been encountered in the field theory dual descriptions of different AdS solutions in recent studies \cite{Faedo:2020lyw,Lozano:2022vsv,Lozano:2024idt}. Closely related to our system in this paper, the $\text{AdS}_3\times \mathbb{CP}^3\times I$ solutions constructed in \cite{Macpherson:2023cbl} were shown to be associated to brane boxes consisting on D3-branes stretched between NS5-branes on one direction and bound states of NS5$'$ and D5-branes at an angle with respect to NS5$'$-branes on a perpendicular direction \cite{Lozano:2024idt}. The 2d gauge theories living in these intersections preserve $(0,3)$ supersymmetries, and were argued to flow in the IR to the 2d $(0,6)$ SCFTs dual to the $\text{AdS}_3\times \mathbb{CP}^3\times I$ solutions.

\begin{figure}[h]
\centering
\includegraphics[scale=0.65]{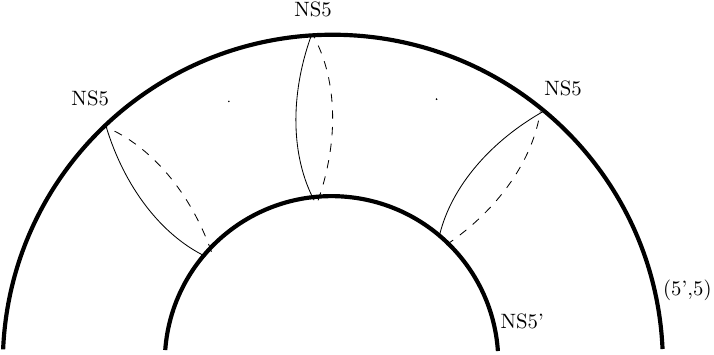}
\caption{Brane set-up associated to Table \ref{ABJM-D2}.}
\label{torus}
\end{figure}

Similarly, in the brane set-up described by Table \ref{ABJM-D2} the D2-branes stretch in $\psi$ between NS5$'$-branes located at $\psi=0$ and $\pi$ and then between $\psi=\pi$ and $\psi=2\pi$. In the $\tau$-direction the D2-brane stretch between NS5-branes located at $\tau_i=\frac{2\pi}{k}i$ and $\tau_{i+1}=\frac{2\pi}{k}(i+1)$, with $i=0,1,\dots,k$. The NS5$'$-brane at $\psi=\pi$ forms a bound state with the orthogonal NS5-brane located at $\tau_i=\frac{2\pi}{k}i$. These bound states are rotated an angle $\phi=\pi/4$ along the 3-7, 4-8 and 7-9 directions with respect to the NS5$'$-branes at $\psi=0,2\pi$.  This is depicted schematically in Figure \ref{torus}. 
The number of D2-branes stretching between the bound state at $\psi=\pi$ and the NS5$'$-brane at $\psi=2\pi$ becomes $k'+p$ because of the fractional D2-branes that arise upon T-dualising the fractional D1-branes found in the Type IIB solution.  The quantum mechanics living on the D2-branes is $\mathcal{N}=3$ supersymmetric, due to the projections induced by the NS5 and the NS5$'$-branes, and the further rotations between the NS5$'$-branes and the (NS5$'$,NS5) bound states. Prior to the rotations the quantum mechanics is $\mathcal{N}=4$ supersymmetric, with SO(4) R-symmetry group realised geometrically as rotations of the $(x^3,x^4,x^5)$ and $(x^7,x^8,x^9)$ directions. When the (NS5$'$,NS5) bound states are rotated the same angle relative to the NS5$'$-branes the R-symmetry reduces to $\text{SO}(3)_R=\text{diag}(\text{SU}(2)\times \text{SU}(2))$, which is the R-symmetry group associated to $\mathcal{N}=3$ supersymmetry, with OSp$(3|2)$ supergroup. As in the ABJM model and the closely related 2d construction in \cite{Lozano:2024idt}, the $\mathcal{N}=3$ quantum mechanics is expected to have the same field content of the $\mathcal{N}=4$ quantum mechanics prior to the rotations of the (NS5$'$,NS5) bound states with respect to the NS5$'$-branes, with the rotations inducing deformations that render some of the multiplets massive. $\mathcal{N}=4$ multiplets in 1d can be obtained by dimensional reduction from 2d $(0,4)$ multiplets  \cite{Okazaki:2015pfa}.  Therefore, one can profit from the extensive literature on 2d $(0,4)$ theories, in particular on their realisations on brane boxes \cite{Hanany:2018hlz}, to construct $\mathcal{N}=4$ quantum mechanics that can later be deformed to $\mathcal{N}=3$.

\begin{figure}[h]
\centering
\includegraphics[scale=0.7]{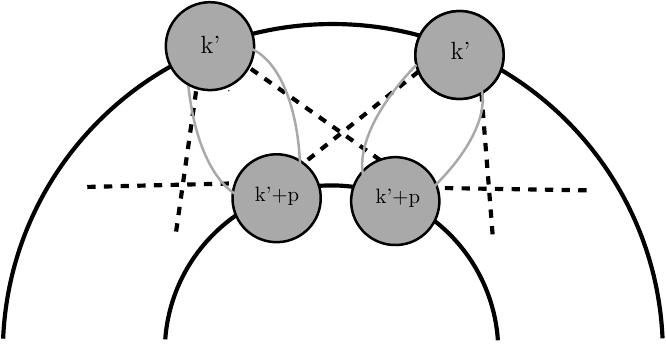}
\caption{Quiver construction associated to the $\text{AdS}_2$ solutions with $h'''=0$. Circles denote $\mathcal{N}=4$ vector multiplets, black lines $\mathcal{N}=4$ twisted hypermultiplets, grey lines $\mathcal{N}=4$ hypermultiplets and black dashed lines $\mathcal{N}=4$ Fermi multiplets. The gauge nodes are repeated $k$ times.} 
\label{quiver-massless}
\end{figure}

The brane box model studied in \cite{Hanany:2018hlz} consists on D3-branes stretched between perpendicular NS5 and NS5$'$-branes. In this brane box a 2d gauge theory with $\mathcal{N}=(0,4)$ supersymmetry and SO(4)$_R$ R-symmetry lives. T-dualising along the common spatial worldvolume direction of this intersection a 1d brane box model is obtained,  that is the one relevant for our analysis. In this brane box D2-branes stretch between NS5 and NS5$'$ orthogonal branes, as depicted in Table \ref{HO}.
\begin{table}[h]
\renewcommand{\arraystretch}{1}
\begin{center}
\scalebox{1}[1]{
\begin{tabular}{c| c cc  c c  c  c c c c}
 branes & $x^0$ & $x^1$ & $\tau$ & $x^3$ & $x^4$ & $x^5$ & $\psi$ & $x^7$ & $x^8$ & $x^9$ \\
\hline \hline
$\mrm{D}2$ & $\times$ & $-$  & $\times$ & $-$ & $-$ & $-$ & $\times$ & $-$ & $-$ & $-$ \\
$\mrm{NS}5'$ & $\times$ & $\times$  & $\times$ & $\times$ & $\times$ & $\times$ & $-$ & $-$ & $-$ & $-$ \\ 
$\mrm{NS}5$ & $\times$  & $\times$ & $-$  & $-$  & $-$  & $-$ & $\times$ & $\times$ & $\times$ & $\times$ \\
\end{tabular}
}
\caption{T-dual of the D3-brane box model studied in \cite{Hanany:2018hlz}, where we have denoted $\tau$ and $\psi$ the field theory directions, as in the brane set-ups associated to our $\text{AdS}_2$ solutions.}
 \label{HO}
\end{center}
\end{table}
Recalling the quantisation of the open strings in the brane box studied in  \cite{Hanany:2018hlz}, we have, dimensionally reducing to 1d,
\begin{itemize}
\item When the end-points of open strings lie on the same stack of D2-branes the projections induced by both the NS5 and NS5$'$-branes leave behind a $\mathcal{N}=4$ vector multiplet.
\item When the end-points of the strings lie on two different stacks of D2-branes separated by an NS5$'$-brane the degrees of freedom along the $(x^7,x^8,x^9)$ directions are fixed, leaving behind the scalars associated to the $(x^3,x^4,x^5)$ directions, which together with the 
$A_\tau$ component of the gauge field give rise to a $\mathcal{N}=4$ twisted hypermultiplet in the bifundamental representation. The hypermultiplets are twisted because the scalars are charged with respect to the SO(3)$_{345}$ subgroup of the R-symmetry group.
\item When the end-points of the strings lie on two different stacks of D2-branes separated by an NS5-brane the degrees of freedom along the $(x^3,x^4,x^5)$ directions are fixed, leaving behind the scalars associated to the $(x^7,x^8,x^9)$ directions, which together with the 
$A_\psi$ component of the gauge field give rise to a $\mathcal{N}=4$ hypermultiplet in the bifundamental representation. The hypermultiplets are untwisted because the scalars are charged with respect to the SO(3)$_{789}$ subgroup of the R-symmetry group.
\item When the end-points of the strings lie on two different stacks of D2-branes separated by both an NS5 and an NS5$'$-brane all the scalars are fixed, leaving behind the fermionic mode associated to a bifundamental $\mathcal{N}=2$ Fermi multiplet.
\end{itemize}
Taking into account this information, the quivers associated to the brane set-up given by Table \ref{ABJM-D2}, underlying the $\text{AdS}_2$ solutions for $h'''=0$, are the ones depicted in Figure \ref{quiver-massless}.
In these quivers circles denote $\mathcal{N}=4$ vector multiplets, black lines $\mathcal{N}=4$ twisted hypermultiplets, grey lines $\mathcal{N}=4$ hypermultiplets and black dashed lines $\mathcal{N}=4$ Fermi multiplets.
Since $\psi$ is a compact direction, there are two $\mathcal{N}=4$ hypermultiplets connecting the $k'$ and $k'+p$ gauge groups at each $\tau$-interval, that originate from the open strings that connect the $k'$ and the $(k'+p)$ D2-branes across either one of the NS5$'$-branes. Similarly, there are two $\mathcal{N}=2$ Fermi multiplets connecting the $k'$ D2-branes with the $(k'+p)$ D2-branes in adjacent $\tau$-intervals, depending on the NS5$'$-brane that is crossed by the open strings. We have denoted these $\mathcal{N}=4$ Fermi-multiplets with black dashed lines. Finally, in each $\tau$-interval the two $\mathcal{N}=4$ bifundamental hypermultiplets connecting the gauge nodes with ranks $k'$ and $k'+p$  combine onto a $\mathcal{N}=6$ bifundamental hypermultiplet. In all, our quiver field theory consists on a sequence of $k$ U$(k')$ and U$(k'+p)$ gauge groups with the field content of $\mathcal{N}=4$ vector multiplets, connected to one another by $\mathcal{N}=6$ bifundamental hypermultiplets. 

It is noteworthy that even if there is no gauge anomaly for the 1d theory, the quiver in Figure \ref{quiver-massless} satisfies the condition for gauge anomaly cancellation for 2d $(0,4)$ gauge theories. This can be checked taking into account the contribution of the different multiplets to the gauge anomaly, summarised in Table \ref{gauge-anomaly} (in 2d $(0,4)$ language). Similar situations have been encountered in other realisations of $\mathcal{N}=4$ supersymmetric quantum mechanics (see \cite{Lozano:2020txg,Lozano:2022vsv}).
\begin{table}[h]
\begin{center}
\begin{tabular}{|c|c|}
\hline
Multiplet & Contribution\\
\hline\hline
(0,4) hyper & $1$ \\
\hline
(0,4) vector  & $-2N$ \\
\hline
(0,2) Fermi& $-\frac12$ \\
\hline
\end{tabular}
\end{center}
\caption{Contribution to the gauge anomaly of the different multiplets that couple in the quivers depicted in Figures \ref{quiver-massless} and \ref{quiver-massive}, in 2d $(0,4)$ language, where $N$ is the rank of the gauge groups.}
\label{gauge-anomaly}
\end{table}

Finally, we would like to emphasise the fact that the quiver depicted in Figure  \ref{quiver-massless} is $\mathcal{N}=4$ supersymmetric, since the effect of the rotations between the bound states of NS5$'$ and NS5-branes at $\psi=\pi$ with respect to the NS5$'$-branes at $\psi=0,2\pi$ has not yet been taken into account. Similarly to what happens in the ABJM theory, the rotations reduce the supersymmetry to $\mathcal{N}=3$, and this translates into some multiplets in the quiver becoming massive. As explained in 
\cite{Lozano:2024idt} for the 2d setting discussed therein, the rotations of the branes mix the twisted and untwisted $(0,4)$ multiplets such that 
$(0,3)$ supersymmetry remains. This applies as well to the 1d case as the 1d multiplets arise upon dimensional reduction from the 2d ones.  This is the analogue in 2d (and in 1d) of the mixing between the $\mathcal{N}=4$ Coulomb and Higgs branches in 3d, resulting in $\mathcal{N}=3$ supersymmetry\footnote{Recall that this means 6 supercharges, and thus, twice the number of supersymmetries of our 1d setting.}.
Unfortunately, since not much is known about $\mathcal{N}=3$ realisations of supersymmetry in 1d (or $(0,3)$ in 2d) we have taken a phenomenological approach. We hope that  our discussion will stimulate a more detailed investigation, that will allow in particular to compute the central charge of the resulting $\mathcal{N}=3$ quivers, that could in turn be compared to the holographic result \eqref{chol}\footnote{With $\text{Vol}(\text{S}^7)$ replaced by $\text{Vol}(\text{S}^7/\mathbb{Z}_k$.)}.

In the next subsection we discuss the field theory associated to the general solutions with non-vanishing Romans mass. Again, we will take a phenomenological approach that will enable us to obtain interesting information, even if partial, about the quantum mechanics dual to the $\text{AdS}_2$ solutions with $\mathcal{N}=6$ supersymmetry.

\subsection{Adding D6-branes}

\begin{figure}[h]
\centering
\includegraphics[scale=0.6]{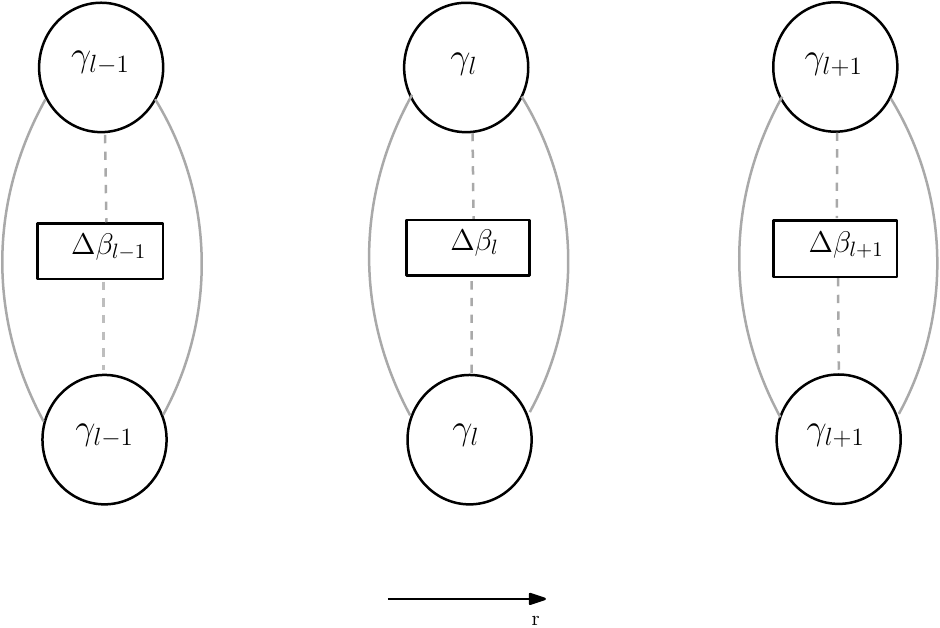}
\caption{Disconnected quivers associated to the general $\text{AdS}_2$ solutions with $\mathcal{N}=6$ supersymmetry along the $r$ direction. $\mathcal{N}=2$ Fermi multiplets have been denoted with grey dashed lines.} 
\label{quiver-massive-r}
\end{figure}

\begin{figure}[h]
\centering
\includegraphics[scale=0.7]{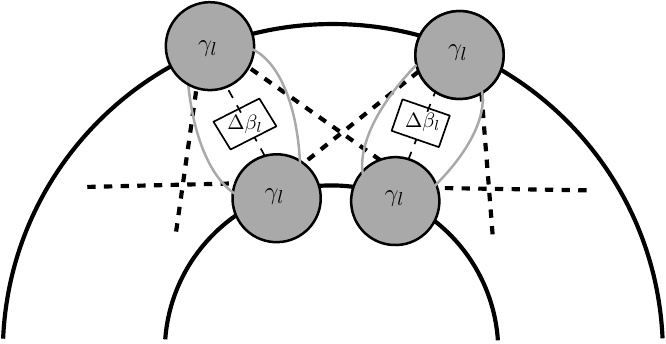}
\caption{Quiver construction associated to the general $\text{AdS}_2$ solutions with $\mathcal{N}=6$ supersymmetry in the $(\tau,\psi)$ torus.}
\label{quiver-massive}
\end{figure}

As we have discussed, the brane intersection associated to the $\text{AdS}_2$ solutions in the general case $h'''\neq 0$ is the one depicted in Table \ref{tablendualised}. It is obtained from the brane set-up depicted in Table \ref{ABJM-D2} adding D6-branes in each $\tau$ interval. Indeed,
the most general solution to the Bianchi identity \eqref{Bianchi} is given by $h_l(r)$ as in \eqref{hder}, from which the numbers of D8 and D0-branes in the original $\text{AdS}_2\times \text{S}^7/\mathbb{Z}_k\times I$ solutions were computed in 
equations \eqref{QD8}, \eqref{QD0}. These become the numbers of D6 and D2-branes in the ``natural'' brane set-up depicted in Table \ref{tablendualised},
\begin{equation}
Q^l_{{\rm D}6}=\beta_l, \qquad Q^l_{{\rm D}2}=\gamma_l,
\end{equation}
where the latter is electric charge, as in equation \eqref{QD0}. Adding the D6-branes into the brane set-up gives rise to new massless modes, associated to the open strings with an end-point on a stack of D2-branes and the other end-point on the stack of D6-branes, in the same $\psi,\tau$ interval. These leave behind a $\mathcal{N}=2$ Fermi multiplet. We thus find a set of disconnected quivers along the $r$-direction similar to the ones encountered in the $\mathcal{N}=8$ case, depicted in Figure 
\ref{D0-F1-D8quiver}, to which we have to add the effect of the $\mathbb{C}^4/\mathbb{Z}_k$ singularity, which introduces  Kaluza--Klein monopoles (or NS5-branes after the T-dualities) as previously discussed. This gives rise to the quivers depicted in Figure \ref{quiver-massive-r}. Again here we assume that the space starts and ends in any of the physically meaningful ways discussed in section 2.2.
In turn, in the $(\tau,\psi)$ torus we find the same quivers depicted in Figure \ref{quiver-massless}, with $\gamma_l$ D2-branes (for $r=l$), no fractional branes, and extra $\mathcal{N}=2$ Fermi multiplets coming from the D6-branes. These quivers are depicted in Figure \ref{quiver-massive}. 

Interestingly, when the $\mathbb{C}^4/\mathbb{Z}_k$ singularity is removed and the $\text{AdS}_2$ solution becomes $\text{AdS}_2\times \text{S}^7\times I$, these quivers become the ones depicted in Figure \ref{D0-F1-D8quiver}. Indeed, in the absence of the KK-monopoles the second family of D2-branes stretched between $\psi=\pi$ and $\psi=2\pi$ disappears. Therefore,  
the two vertical $\mathcal{N}=4$ hypermultiplets depicted in Figure \ref{quiver-massive-r} become adjoint hypermultiplets. These produce three $\mathcal{N}=4$ adjoint hypermultiplets, and therefore an adjoint $\mathcal{N}=8$ hypermultiplet, together with the hypermultiplet that connects adjacent $\gamma_k$ nodes along the $\tau$-direction (see Figure \ref{quiver-massive})\footnote{See \cite{Garcia-Compean:1998sla} for the decomposition of $\mathcal{N}=8$ multiplets into $\mathcal{N}=4$ ones, in 2d language.}, which becomes an adjoint hypermultiplet when $k=1$. In turn, the $\mathcal{N}=4$ Fermi multiplets that connect diagonally the $\gamma_k$ node with the adjacent ones in  Figure \ref{quiver-massive} produce an adjoint $\mathcal{N}=4$ Fermi multiplet. This gives rise to a $\mathcal{N}=8$ vector multiplet together with the $\mathcal{N}=4$ vector multiplet associated to the $\gamma_k$ node (see \cite{Garcia-Compean:1998sla}). In this way, the $\mathcal{N}=8$ quivers depicted in Figure \ref{D0-F1-D8quiver}, associated to the $\text{AdS}_2\times \text{S}^7\times I$ solutions, are nicely recovered.

Analogously to our discussion in subsection \ref{sec:N=8FT}, in the $h'''\neq 0$ (massive) case the D2-branes stretched along the $\tau$ and $\psi$ directions can be shown to play the role of baryon vertices for the D6-branes lying in these intervals, given the conditions \eqref{betagamma}. The relative orientations between the D2-F1-D6 branes in Table \ref{tablendualised} are indeed the ones that allow to create F1-strings stretched between the D2 and the D6-branes. The coupling in the WZ action of the D2-branes that allows for this is
\begin{equation}\label{Wilson-couplings}
S_{{\rm D}2}=T_{2}\int_{t,\tau,\psi}\hat{F}_{2}\wedge A_t=T_{F1}Q_{{\rm D}6}\int A_t\, .
\end{equation}

The interpretation of the $\text{AdS}_2\times \text{S}^7/\mathbb{Z}_k\times I$ solutions in the massive case is thus as describing backreacted baryon vertices in the field theory living in D8-branes probing a $\mathbb{C}^4/\mathbb{Z}_k$ singularity. The D0 baryon vertices break the 12 supersymmetries preserved by the D8-branes by a half, resulting in 1d $\mathcal{N}=6$ supersymmetry.  As for the $\text{AdS}_2\times \text{S}^7\times I$ solutions, $\mathcal{N}=8$ supersymmetric, the theory living in the D8-branes on $\mathbb{C}^4/\mathbb{Z}_k$  is not superconformal, but becomes a superconformal quantum mechanics when the D0-branes and the F1-strings are added to the configuration.

\section{Conclusions}
In this work, we considered a class of local solutions in massive Type IIA that are a warped product of AdS$_2\times$S$^7$ and an interval $I$, constructed in \cite{Dibitetto:2018gbk}. These solutions have OSp$(8|2)$ symmetry and are charaterised by a semi-infinite interval which can be bounded at one end by one of a regular space, OF1-plane or D8/O8 system. We generalise these solutions by taking a $\mathbb{Z}_k$ orbifold of the seven-sphere expressed as a U(1) fibration over $\mathbb{CP}^3$, which breaks supersymmetry to ${\cal N}=6$ and the superconformal symmetry to OSp$(6|2)$. We further establish that it is possible to place source D8-branes along the interior of $I$, which can be used to glue local solutions together, allowing one to construct compact solutions.

The above ${\cal N}=6$ solution permits one to T-dualise on the Hopf fiber of U(1) $\hookrightarrow$ S$^7\to \mathbb{CP}^3$ resulting in new solutions of the form AdS$_2\times\mathbb{CP}^3\times$S$^1\times I$ in Type IIB. In the limit of vanishing axion, we are able to establish a web of dualities relating our solutions both to another class of $\text{AdS}_2\times \mathbb{CP}^3\times \text{S}^1\times I$ solutions constructed in \cite{Conti:2023rul} and to AdS$_4/\mathbb{Z}_{k'}\times$S$^7/\mathbb{Z}_k$ in eleven dimensions. This aids us in determining the brane intersections realising our ${\cal N}=8$ and ${\cal N}=6$ solutions in general.

With the brane intersections in hand, we make proposals for the SCQM that our ${\cal N}=8$ and ${\cal N}=6$ solutions are dual to, in terms of disconnected quivers that should be realised as the IR fixed point of an RG flow experienced by the theory living on a D0-F1-D8 (plus KK-monopole for ${\cal N}=6$) system.  We also give an interpretation of our geometries as describing baryon vertices. \\

Considered in tandem, the Type IIB ${\cal N}=6$ solution we generate by T-dualising on the Hopf fiber of U(1) $\hookrightarrow$ S$^7\to \mathbb{CP}^3$ and the solutions of \cite{Conti:2023rul}, generated by AdS$_3$ Hopf fiber T-duality of ${\cal N}=(6,0)$ AdS$_3$ solutions in massive Type IIA, \cite{Macpherson:2023cbl} are interesting. They  both yield an AdS$_2$ solution defined by an order 3 polynomial, but while they are related by S-duality for vanishing axion, they are not in general related to each other by any dualities. This suggests a more general underlying class of ${\cal N}=6$ solutions that are foliations of AdS$_2\times \mathbb{CP}^3$ over a Riemann surface; this class does indeed exist and will be reported in \cite{Conti:2025djz}.

Owing to a lack of development of SCQM realising OSp$(6|2)$ and OSp$(8|2)$ superconformal symmetries we have been forced to take a somewhat phenomenological approach when proposing  the holographic duals to our AdS$_2$ solutions. This lack of development is quite evident if one attempts to compute the central charge of the dual SCQM, whose holographic analogue can be quite easily derived from the Brown--Henneaux formula \cite{Brown:1986nw}. Even if our understanding of their dual quantum mechanics is only partial, our ${\cal N}=8$ and ${\cal N}=6$  AdS$_2$ solutions are well-defined solutions in supergravity and should have a well-defined dual. Given the high amount of supersymmetry involved, it should be possible to precisely establish this duality. It is our hope that our work will motivate a more detailed study in this direction in the future.

Additional future directions include:

\begin{itemize}

\item It should be possible to extend the class of $\text{AdS}_2$ solutions in Type IIA to also include fractional branes. The reason why we do not see these charges in the massless case, but we do see them in Type IIB after T-duality, is that we need to use fluxes and potentials that do explicitly take into account that there is an isometric direction. If we used these fluxes we would be able to see the D0 fractional charges in the original formulation in Type IIA. The most general solutions in Type IIB do however contain these couplings, because the T-duality ``brings them to light''.  

\item After a T-duality along the Hopf fibre of the $\mathbb{CP}^3$ a new class in Type IIA arises consisting of $\text{AdS}_2\times \text{S}^2\times \text{S}^2\times \text{S}^1\times \text{S}^1$ fibration over a Riemann surface, preserving $\mathcal{N}=4$ supersymmetries. This is suggestive of a more general class of solutions preserving OSp$(4|2)$ superconformal symmetry which may admit an enhancement to large ${\cal N}=4$. Such a class should have interesting applications in the context of black hole near-horizons. 

\item The connection with the $\text{AdS}_2\times \mathbb{CP}^3\times \text{S}^1\times I$ solutions constructed in \cite{Conti:2023rul} allows us to extend the class of $\text{AdS}_2\times \text{S}^7\times I$ solutions in \cite{Dibitetto:2018gbk} to a more general class, in which the S$^7$ is deformed to preserve $\mathcal{N}=5$ supersymmetries. The key ingredient here is that in \cite{Conti:2023rul} a further class of solutions where the $\mathbb{CP}^3$ was deformed to preserve $\mathcal{N}=5$ supersymmetry was constructed, and we can S- plus T-dualise the solution corresponding to vanishing axion field to generate a massless, deformed, $\text{AdS}_2\times \tilde{\text{S}}^7\times I$ solution with $\mathcal{N}=5$ supersymmetry, that we can then extend to the massive case . Along the way, the massless, deformed solution can be uplifted to eleven dimensions to produce a new $\text{AdS}_3 \times \tilde{\text{S}}^7\times I$ solution to M-theory with $\mathcal{N}=5$ supersymmetry.

\item As shown in \cite{Legramandi:2023fjr} it is possible to replace S$^7/\mathbb{Z}_k$ by any weak G$_2$-manifold.

\end{itemize}

\section*{Acknowledgements}
NTM and YL thank Andrea Conti for collaboration on related topics and acknowledge support from grants from the Spanish government MCIU-22-PID2021-123021NB-I00 and principality of Asturias SV-PA-21-AYUD/2021/52177. NTM is also supported by the Ram\'on y Cajal fellowship RYC2021-033794-I.

\end{document}